\documentclass[aps,prb,twocolumn,secnumarabic,balancelastpage,amsmath,amssymb,nofootinbib,floatfix,superscriptaddress,10pt]{revtex4-2}
\usepackage{graphicx}      % tools for importing graphics
\usepackage{siunitx}
\usepackage{soul}
\usepackage{braket}
\usepackage{float}
\usepackage{booktabs,caption}
\usepackage{array}
\usepackage[flushleft]{threeparttable}
\usepackage{textcomp}
\usepackage{adjustbox}
\usepackage{tabularx}
\usepackage{xcolor,colortbl}
\definecolor{Gray}{gray}{0.85}
\definecolor{LightCyan}{rgb}{0.88,1,1}
\DeclareUnicodeCharacter{2062}{\,}
\usepackage[colorlinks=true]{hyperref}  % this package should be added after 
\begin{document}
\raggedbottom

\title{Importance of anisotropic interactions for hard-axis or hard-plane ordering of Ce-based ferromagnets}
\author{Hanshang Jin}
\affiliation{Department of Physics and Astronomy, University of California, Davis, California 95616, U.S.A.}
\author{Rahim R. Ullah}
\affiliation{Department of Physics and Astronomy, University of California, Davis, California 95616, U.S.A.}
\author{Peter Klavins}
\affiliation{Department of Physics and Astronomy, University of California, Davis, California 95616, U.S.A.}
\author{Valentin Taufour}
\affiliation{Department of Physics and Astronomy, University of California, Davis, California 95616, U.S.A.}

\begin{abstract}
Ferromagnetic (FM) Kondo-lattice (KL) compounds often exhibit intriguing magnetic behavior driven by strong crystal electric field (CEF) anisotropy and exchange interactions. Recent studies suggest that many Ce- and Yb-based KL ferromagnets order along the hard-axis or hard-plane determined by their CEF ground state. We performed a survey of Ce-based FM compounds, complemented with new single crystal synthesis, magnetization measurements, and analysis of the CEF scheme. Our results reveal that hard-axis or hard-plane ordering is less common than previously reported, with most compounds ordering along their easy-axis or easy-plane. We also find no clear correlation between the strength of the Kondo effect and whether a compound adopts hard-axis or hard-plane or easy-axis or easy-plane ordering. Instead, the key driver appears to be the competition between CEF anisotropy and exchange interaction anisotropy. A direct comparison of CeCuSi and CeAgSb$_2$ further indicates that antiferromagnetic interactions along the CEF easy-axis can be crucial in stabilizing the hard-axis FM ordering observed in CeAgSb$_2$. Finally, our results suggest that an anisotropic Ruderman–Kittel–Kasuya–Yosida model could offer deeper insights into the complicated magnetic properties of Ce-based and other heavy-fermion systems.
\end{abstract}

\maketitle

%%%%%%%%%%%%%%%%%%%%%%%%%%%%%%%%%%%%%%%%%%%%%%%%%%%%%%%%%%%%%%%%%%

\section{Introduction}
The study of Kondo lattice (KL) systems has attracted significant interest over the past few decades due to their complex interactions and exotic ground states ~\cite{stewart1984Heavyfermion,stewart2001NonFermiliquid,brando2016Metallic,weng2016Multiple,yamamoto2010Metallic}. In many KL systems, the localized magnetic ions are subject to the crystal electric field (CEF) effect, which arises from the electrostatic interaction between the magnetic ions and their surrounding ions. The CEF splits the degenerate energy levels of the magnetic ions which leads to anisotropic magnetic properties. 

One intriguing phenomenon observed in some KL systems is the ferromagnetic (FM) ordering along the CEF hard axis or hard plane, contrary to the expected alignment along the easy axis or easy plane~\cite{hafner2019Kondolattice}. Magnetic hard-axis ordering is believed to result from a strong exchange anisotropy, which eventually surpasses the CEF anisotropy~\cite{araki2003Crystal, nikitin2021Magnetic, hafner2019Kondolattice}. Another proposed explanation for magnetic hard-axis ordering involves the substantial influence of strong Kondo fluctuations~\cite{kruger2014FluctuationDriven,kwasigroch2022Magnetic}.

We surveyed Ce-based FM KLs with available single-crystal magnetization data, and performed additional analysis to determine the CEF scheme. Our results show that the majority of Ce-based FM KLs are ordered along the CEF easy-axis, contrary to the previous report~\cite{hafner2019Kondolattice}. Only 5 out of 30 of these compounds show hard-axis or hard-plane ordering. Three of these 5 have a relatively larger difference between in-plane and out-of-plane Ce-Ce distance($>$ 1.9\,\AA), and 2 of these 5 have more than one Ce site. Our survey suggests that hard-axis ordering is mainly due to the competition between CEF anisotropy and exchange interaction anisotropy. The comparison of CeCuSi (easy-axis) and CeAgSb$_2$ (hard-axis) provides a challenge to the importance of the Kondo effect: CeCuSi exhibits a stronger Kondo effect than CeAgSb$_2$ but without hard-axis ordering. Our CEF analysis suggests that CeAgSb$_2$ exhibits an AFM interaction along the $ab$ plane that could lead to the moments favoring the CEF hard axis. Furthermore, our findings suggest that it is important to consider an anisotropic Ruderman–Kittel–Kasuya–Yosida (RKKY) model when analyzing Ce-based and other heavy fermion systems to better understand their complex magnetic properties.
%%%%%%%%%%%%%%%%%%%%%%%%%%%%%%%%
%%%%%%%%%%%%%%%%%%%%%%%%%%%%%%%
\section{Methods}

The CEF plays a crucial role in determining the magnetic properties of $4f$ materials. The inverse magnetic susceptibility in heavy fermion compounds often deviates from the Curie-Weiss law and is mainly due to the CEF effect~\cite{banda2018Crystalline,kabeya2022Eigenstate,ajeesh2017Isingtype}. The theoretical expression of the magnetic susceptibility with the Van Vleck contribution is given by
\begin{equation} \label{eq:chi}
\begin{aligned}
\chi_{i} =\frac{N_{A}g_{J}^{2}\mu _\mathrm{{B}}^{2}\mu_{0} }{Z}&\Big[\sum_{n}^{}\beta|\braket{J_{i,n}}|^2e^{-\beta \Delta_n}\\ + 2&\sum_{m\neq n}|\Braket{m|J_{i,n}|n}|^2\left(\frac{e^{-\beta \Delta_{m}}-e^{-\beta \Delta_{n}}}{\Delta_{n}-\Delta_{m}}\right)\Big],
\end{aligned}
\end{equation}
where $\beta=1/k_BT$, $Z=\sum_{n}e^{-\beta \Delta_{n}}$, $i=x,y,z$, and $n,m=0,1,2$ for the three Kramers doublets of Ce$^{3+}$ with $J=5/2$ under a noncubic symmetry.

Depending on the site symmetry of the Ce ions in the crystal, we can determine the corresponding CEF Hamiltonian with the CEF parameters $B_m^n$ and the Stevens operators $O_m^n$~\cite{stevens1952Matrix,hutchings1964Pointcharge}. For Ce ions, it is possible to derive the analytical expressions of magnetic susceptibility along various directions exhibiting higher symmetry, such as tetragonal, trigonal, and hexagonal symmetry. The exact analytical expressions for tetragonal symmetry are provided in Appendix~\ref{CeAgSb2}, and those for trigonal and hexagonal symmetry can be found in Refs.~\cite{jin2024Easyplane,jin2022Suppression}, respectively.

The variables of the analytical expressions of magnetic susceptibility along various directions are the CEF parameters $B_m^n$. These parameters can be simultaneously fitted from the experimental magnetic susceptibilities measured along different directions. The first CEF parameter $B_{2}^{0}$ can be estimated from the paramagnetic Curie-Weiss temperatures $\theta_{CW}^{\parallel}$ and $\theta_{CW}^{\perp}$ by the following expression~\cite{bowden1971Crystal,wang1971Crystalfield}, which gives the strength of the magnetocrystalline anisotropy:
\begin{equation} \label{B20CW}
B_{2}^{0} = (\theta _{CW}^{\perp}-\theta _{CW}^{\parallel})\frac{10k_{B}}{3(2J-1)(2J+3)}.
\end{equation}
A negative $B_{2}^{0}$ means that ground-state moment is primarily aligned along the $c$ axis, and a positive $B_{2}^{0}$ means that moments favor $ab$ plane. The estimated $B_{2}^{0}$ can be used as a starting value, before being released during fitting.

The inverse magnetic susceptibility including the molecular field contribution $\lambda_{i}$ and the residual susceptibility $\chi_{0}^{i}$ is calculated as~\cite{kabeya2022Eigenstate}:
\begin{equation} \label{eq:inverchi}
\chi_{i}^{-1}=\left ( \frac{\chi_{i}^\mathrm{CEF}(T)}{1-\lambda_{i} \chi_{i}^\mathrm{CEF}(T)} + \chi_{0}^{i} \right )^{-1}.
\end{equation}
The sign of the molecular field contribution $\lambda_{i}$ reflects the nature of the exchange interactions. A positive value indicates an overall FM exchange interaction, while a negative value suggests an overall AFM exchange interaction. The RKKY interaction can be either FM or AFM, depending on the distance between magnetic ions and the characteristics of the Fermi surface. The Kondo screening effect is always AFM. Therefore, interpreting the value of $\lambda_{i}$ requires considering contributions from both RKKY and Kondo interactions.

A reliable CEF scheme should be consistent with all of the physical properties related to the CEF ground state and the splitting of CEF energies. The CEF ground state influences the preferred magnetic moment direction at low temperatures, as well as the saturation moment along different axes. The splitting of CEF energies affects the temperature dependence and anisotropy of the magnetic susceptibility, often leading to a characteristic curvature that deviates from the Curie-Weiss law. Additionally, CEF splitting gives rise to Schottky anomalies in heat capacity and influences electron scattering processes, though isolating its effects from other scattering mechanisms can be challenging. Inelastic neutron scattering and Raman spectroscopy are also useful methods to determine the CEF splitting and the symmetry of the ground states~\cite{zhang2021Crystalline, cai2024Crystalline}.

Therefore, to obtain accurate CEF parameters and molecular field contributions, it is necessary to apply constraints from the ground state and the splitting energies if these are available. Often, we use the Schottky anomaly to determine the splitting energies. For a three-level system, the Schottky anomaly can be expressed as the following expression~\cite{souza2016Specific}:
\begin{equation} \label{HCSCH}
\begin{split}
C_{sch} =\frac{R}{(k_BT)^2}\frac{e^{(\Delta_1+\Delta_2)/k_BT}}{(e^{\Delta_1/k_BT}+e^{\Delta_2/k_BT}+e^{(\Delta_1+\Delta_2)/k_BT})^2}\\
\times[-2\Delta_1\Delta_2+\Delta_2^2(1+e^{\Delta_1/k_BT})+\Delta_1^2(1+e^{\Delta_2/k_BT})]
\end{split}
\end{equation}

Single-crystal magnetic susceptibility data are crucial for determining the CEF scheme. Single crystals of CeTiGe$_3$~\cite{jin2022Suppression}, CeCuSi~\cite{jin2024Easyplane}, and CeAgSb$_2$~[\ref{CeAgSb2}] have been synthesized using flux growth. Magnetic properties were measured using a SQUID magnetometer (Quantum Design MPMS XL) with 1\,T magnetic field, in the temperature range of 2$-$300\,K.

In addition to these three samples, we have also performed detailed CEF analysis on five other samples that had not been previously analyzed. The data were digitized from the online publications using PlotDigitizer~\cite{2024PlotDigitizer}. Depending on the quality of the image and the algorithm limits in this software, the digitized data could differ slightly from the original data. Some of these analysis yield reasonably accurate results, which are shown in Figs~\ref{CeFM_tetra_figs} and \ref{CeFM_tri_figs}, while others do not. Detailed CEF analyses and discussions for all the compounds are provided in the Appendix. We also discuss the limitations of the CEF analysis in Appendix~\ref{limitation}.

%%%%%%%%%%%%%%%%%%%%%%%%%%%%
%%%%%%%%%%%%%%%%%%%%%%%%%%%%
%%%%%%%%%%%%%%%%%%%%%%%%%%%

%%%%%%%%%%%%%%
%%%%%%%%%%%%%%

%%%%%%%%%%%%%%%%%%%%%%%%%%%%%%%%%%%%%%%%%%%%%%%%
%%%%%%%%%%%%%%%%%%%%%%%%%%%%%%%%%%%%%%%%%%%%%%%%
%%%%%%%%%%%%%%%%%%%%%%%%%%%%%%%%%%%%%%%%%%%%%%%%
%%%XXX%%%%XXXXX%%%XXX%%%%XXXXX
%%%XXX%%%%XXXXX%%%XXX%%%%XXXXX
%%%%%%%%%%%%%%%%%%%%%%%%%%%%%%

\begin{table*}[!htb]
\center
\caption[List of Ce-based ferromagnetic Kondo Lattice compounds]{List of Ce-based FM KL compounds that have available single crystal magnetization data, with selected physical properties.   $d^\parallel_{\textrm{Ce}\mbox{-}\textrm{Ce}}$ represents in-plane Ce-Ce distance, and $d^\perp_{\textrm{Ce}\mbox{-}\textrm{Ce}}$ represents out-of-plane Ce-Ce distance, and their differences are denoted $\Delta d_{\textrm{Ce}\mbox{-}\textrm{Ce}} = | d^\perp_{\textrm{Ce}\mbox{-}\textrm{Ce}} - d^\parallel_{\textrm{Ce}\mbox{-}\textrm{Ce}}|$. CEF ground-state anisotropy = $\Braket{J_x}$/$\Braket{J_z}$ ($>$1 favors $ab$ plane, $<$1 favors $c$ axis).  $*$ = value obtained based on our measurements. $\dag$ = value obtained based on published data. H$_{cr}$ represents the field at which the $M(H)$ curves along the hard and easy-axis cross each other at 2\,K. The detailed discussion can be found in Sec.~\ref{CeFM}.}
\begin{tabular}{|p{0.19\linewidth}|p{0.04\linewidth}|p{0.08\linewidth}|p{0.078\linewidth}|p{0.05\linewidth}|p{0.05\linewidth}|p{0.07\linewidth}|p{0.07\linewidth}|p{0.055\linewidth}|p{0.055\linewidth}|p{0.058\linewidth}|p{0.095\linewidth}|}
\bottomrule
Compound & $T_\textrm{C}$ (K)& $S_{4f}/R\ln2$ at $T_\textrm{C}$ & Ce point group\footnotemark[1] & $d^\perp_{\textrm{Ce}\mbox{-}\textrm{Ce}}$ (\AA)& $d^\parallel_{\textrm{Ce}\mbox{-}\textrm{Ce}}$ (\AA) & $\Delta d_{\textrm{Ce}\mbox{-}\textrm{Ce}}$ (\AA) & Ordering axis & CEF ground state ani\-so\-tro\-py & $\lambda_{\parallel}$ (mol/ e.m.u.)& $\lambda_{\perp}$ (mol/ e.m.u.) & $M_\parallel$, $M_\perp$ crossing    \\ \hline

\rowcolor{LightCyan}\multicolumn{12}{ |c| }{\textbf{Case I: Easy-axis ordering}} \\ \hline
CeTiGe$_3$~\cite{jin2022Suppression,inamdar2014Anisotropic},[\ref{CeTiGe3_appdix}] & 14.0 & 1.13\footnotemark[6] & $D_{3h}$ & 6.271   & 4.664 & 1.607 & $c$ axis & 0$^*$ & 2.2$^*$ & -50.3$^*$&  No crossing\\

CeGaGe~\cite{ram2023Magnetic},[\ref{CeGaGe_appdix}] & 6.0 & 0.76 & $C_{2v}$ & 4.292   & 4.217 & 0.075 & $c$ axis &0.42& 10 & 18.7  & No crossing\\

CeRu$_2$Al$_2$B\footnotemark[2]\cite{matsuoka2013Magnetic,matsuoka2012Ferromagnetic,baumbach2012CeRu2Al2B},[\ref{CeRu2Al2B_appdix}] & 13 & 0.75 & $D_{4h}$ & 4.198   & 5.607 & 1.409 &$c$ axis&0.31& 3.06$^\dag$   & -36.9$^\dag$  &No crossing\\ 
CePd$_2$P$_2$~\cite{drachuck2016Magnetization,jeitschko1983Ternary,tran2014Ferromagnetism},[\ref{CePd2P2_appdix}] & 28.4 & 0.64 & $D_{4h}$ & 4.156   & 5.751 & 1.595 &$c$ axis& 0.15 & 18.9$^\dag$   &  59.5$^\dag$   &No crossing\\
Ce$_3$Al$_{11}$\footnotemark[3]\cite{s.garde2008Electrical},[\ref{Ce3Al11_appdix}] & 6.3 & 0.63 & $D_{2h}$,$C_{2v}$ & 4.260    & 5.999  & 1.739  & $b$ axis & 1.3 & -9 & -5, -20& No crossing\\
CeVSb$_3$~\cite{sefat2008Magnetization},[\ref{otherCe_appdix}] & 4.6 & 0.54 & $C_s$ & 4.322   & 5.840 & 1.519 &$c$ axis&&   &     &No crossing\\
Ce$_7$Rh$_3$~\cite{tsutaoka2008Ferromagnetic,sereni1992Coexistence}[\ref{otherCe_appdix}] & 6.8 & 0.70 & $C_{s}$,$C_{s}$,$C_{3v}$ & 4.576    & 3.821  & 0.756  & $c$ axis &  &  & &  No crossing\\

CePtAl~\cite{kitazawa1997Magnetic,donni1995Commensurateincommensurate},[\ref{otherCe_appdix}] & 5.9 & 0.72 & $C_{s}$ & 3.753    & 4.482  & 0.729  & $a$ axis &  &  & & No crossing\\

Ce$_2$Pd$_2$In\footnotemark[3]\cite{klicpera2016Magnetic}[\ref{otherCe_appdix}] & 4.1 & 0.66 & $D_{2v}$ & 4.075   & 3.916 & 0.159 & $c$ axis &  &  & &  No crossing\\

\hline\rowcolor{LightCyan}\multicolumn{12}{ |c| }{\textbf{Case II: Easy-plane ordering}} \\ \hline
CeCuSi~\cite{jin2024Easyplane} & 15.5 & 0.81$^*$ & $D_{3d}$ & 4.239   & 3.986 & 0.253 & $ab$ plane & 4.6$^*$ & 9$^*$ & 5.2$^*$ &  No crossing \\
CeNiSb$_2$~\cite{thamizhavel2003Anisotropic},[\ref{CeNiSb2_appdix}] &  6.0 & 0.52 & $C_{4v}$ & 4.409   & 5.800 & 1.391 & $ab$ plane & 2.7 & 19 & 19   & No crossing\\
CePdSb~\cite{katoh1999Anisotropic},[\ref{CePdSb_appdix}] & 17.5 & 0.97 & $D_{3d}$ & 4.598   & 3.957 & 0.641 & $ab$ plane & 4.8 & 35.0$^\dag$ & 14.4$^\dag$ &  No crossing\\

CePtSb~\cite{katoh1999Anisotropic},[\ref{CePtSb_appdix}] & 4.7 & 0.76 & $D_{3d}$ & 4.543   & 4.042 & 0.502 & $ab$ plane & 6.9 & 45.6$^\dag$ & 4.10$^\dag$ &  No crossing\\

%\hline\rowcolor{LightCyan}\multicolumn{12}{ |c| }{\textbf{ Uncategorized Easy-plane ordering cases }} \\ 
\hline
CeRh$_6$Ge$_4$~\cite{shu2021Magnetic,matsuoka2015Ferromagnetic,shen2020Strangemetal,vosswinkel2012Bismuth},[\ref{CeRh6Ge4_appendix}]  & 2.5 & 0.19 & $D_{3h}$ & 7.154   & 3.855 & 3.299 & $ab$ plane & 3.0 & -111?\footnotemark[9] & -52.8?\footnotemark[9] &No crossing\\
CeCuGa$_3$~\cite{joshi2012Magnetic},[\ref{CeCuGa3_appendix}] & 4.2 & 0.71 & $D_{4h}$ & 4.266   & 6.026 & 1.760 & $ab$ plane & 3.0 &  &   &No crossing\\
CeNi$_{0.74}$Ga$_{3.26}$~\cite{haddock2022Flux},[\ref{CeNiGa3_appendix}] & 5.6 & 0.75$^\dag$ & $D_{4h}$ & 4.316   & 6.007 & 1.691 &$ab$ plane&3.0$^\dag$ & &    &No crossing\\
CeAuGa$_3$~\cite{lv2022Transport},[\ref{otherCe_appdix}] & 1.8 & 0.53 & $D_{4h}$ & 4.340   & 6.150 & 1.810 & $ab$ plane &3.0$^\dag$ &  &    &No crossing\\ 
CeGa$_2$\footnotemark[4]\cite{takahashi1988Multiple},[\ref{otherCe_appdix}] & 8.4 & 0.75 & $D_{6h}$ & 4.283   & 4.332 & 0.050 & $ab$ plane & 3.0 &  &  &  No crossing\\  \hline
CePtAl$_4$Si$_2$~\cite{ghimire2014Investigation},[\ref{otherCe_appdix}] & 3.7 & 0.48 & $D_{4h}$ & 4.271   & 8.060 & 3.789 & $ab$ plane &   &   &   &  No crossing\\
CeCr$_2$Si$_2$C~\cite{wang2023Observation},[\ref{otherCe_appdix}] & 12.4 & 0.08 & $D_{4h}$ & 4.013   & 5.278 & 1.265 & $ab$ plane & &  &   &No crossing\\
CeAlSi~\cite{yang2021Noncollinear,wang2021Structure},[\ref{otherCe_appdix}] & 8.2 & 0.83 & $C_{2v}$ & 4.250   & 4.219 & 0.032 & $ab$ plane & &   &    &No crossing\\
CeAgAl$_3$~\cite{muranaka2007Thermodynamic,nallamuthu2017Ferromagnetism},[\ref{otherCe_appdix}] & 3.0 & 0.60 & $C_{2v}$ & 4.358  & 6.138 & 1.780 &$ab$ plane&&   &     &No crossing\\ 
Ce$_5$Pb$_3$O~\cite{macaluso2004Structure}[\ref{otherCe_appdix}] & 46 & - & $C_{s}$,$C_{4h}$ & 4.161   & 3.834 & 0.327 & $ab$ plane &  &  & &  No crossing\\

CeRh$_3$B$_2$~\cite{dhar1981Strong,galatanu2003Unusual}[\ref{otherCe_appdix}] & 115 & - & $D_{6h}$ & 5.740   & 3.080 & 2.390 & $ab$ plane &  &  & &  No crossing\\

CeIr$_3$B$_2$~\cite{kubota2013Weak}[\ref{otherCe_appdix}] & 41 & - & $C_{2h}$ &    5.500 &3.090 &2.410 & $ab$ plane &  &  & &  No crossing\\

\hline\rowcolor{LightCyan}\multicolumn{12}{ |c| }{\textbf{Case III: Hard-axis Ordering, CEF favors $ab$ plane}} \\ \hline
CeAgSb$_2$~\cite{araki2003Crystal,nikitin2021Magnetic},[\ref{CeAgSb2}] & 9.6 & 0.9 & $C_{4v}$ & 4.368   & 6.390 & 2.023 &$c$ axis& 3.0 & 12.07$^*$& -15.06$^*$   & H$_{cr}$=1\,T\\
CeRuPO~\cite{krellner2008Single,krellner2007CeRuPO},[\ref{CeRuPO_appdix}] & 15 & 0.69 & $C_{4v}$ & 4.026   & 6.484 & 2.458 &$c$ axis&3.0& 33.1$^\dag$  &  -25.3$^\dag$   & H$_{cr}\approx$0.3\,T\\
CeFeAs$_{0.7}$P$_{0.3}$O~\cite{jesche2012Ferromagnetism},[\ref{otherCe_appdix}] & 7.5 & - & $C_{4v}$ & 4.006  & 6.811 & 2.806 &$c$ axis&3.0$^\dag$ & & &H$_{cr}\approx$0.3\,T\\
$\beta$-CeNiSb$_3$~\cite{thomas2007Discovery},[\ref{otherCe_appdix}] & 6 & 0.70 & $C_2,C_{s}$ & 4.302   & 5.937 & 1.634 &$c$ axis\footnotemark[8]&&  & &H$_{cr}\approx$0.7\,T\\

\hline\rowcolor{LightCyan}\multicolumn{12}{ |c| }{\textbf{Case IV: Hard-plane Ordering, CEF favors $c$ axis}} \\ \hline
Ce$_{2}$PdGe$_{3}$ \footnotemark[5]\cite{baumbach2015Complex,bhattacharyya2016Exploring} & 2.25 & 0.16\footnotemark[7] & $D_{2h},D_{2d}$ & 4.244  & 4.264 & 0.020 & $ab$ plane &&  &  &H$_{cr}\approx$1\,T\\ \bottomrule
 
\end{tabular}

\footnotetext[1]{all Ce sites point group were listed if there is more than one Ce site}
\footnotetext[2]{AFM transition along both axes at $T_\textrm{N}$=14.3\,K}
\footnotetext[3]{$T_N$ = 4.5\,K}
\footnotetext[4]{$T_{N1}$ = 9.9\,K, $T_{N1}$ = 10.3\,K, $T_{N3}$ = 11.3\,K}
\footnotetext[5]{AFM transition only along $c$ axis at $T_{\textrm{N}_1}=10.7$\,K and $T_{\textrm{N}_2}=9.6$\,K}
\footnotetext[6]{entropy reaches to $R/\ln(2)$ near $12.9$\,K}
\footnotetext[7]{entropy reaches to 0.77$R/\ln(2)$ near $T_{\textrm{N}_1}=10.7$\,K}
\footnotetext[8]{CEF favors $b$ aixs, then $c$ axis, and lastly $a$ axis. The ground state favors $c$ axis, then $b$ axis, and lastly $a$ axis.}
\footnotetext[9]{see discussion in main text and Appendix~\ref{CeRh6Ge4_appendix}}
\label{CeFMKLs}
\end{table*}

\section{Results}\label{CeFM}
To investigate the hard-axis ordering phenomenon, we conducted an extensive survey of Ce-based FM KLs with available single-crystal magnetic susceptibility data. Compounds with cubic symmetry and those containing other magnetic elements such as Fe, Co, or Mn were excluded from this study. Compared to the Ce-based compounds mentioned in Ref.~\cite{hafner2019Kondolattice}, we found 22 additional Ce compounds that have available single-crystal magnetization data. %We remove CeIrGe$_3$~\cite{anand2018Understanding} from the list~\cite{hafner2019Kondolattice} since all three transitions are AFM and there is no clear evidence to indicate whether CeIrGe$_3$ exhibits hard-axis ordering or not~\cite{kawai2008Magnetic}.

We compile the transition temperature $T_\textrm{C}$, the entropy $S_{4f}/R\ln2$ at $T_\textrm{C}$ to estimate the Kondo effect strength, their Ce point group, and their ordering axis. Based on the published single-crystal magnetization measurements, we can also report whether there is a magnetization crossing or not. For the hard-axis or hard-plane ordering, the crossing fields, H$_{cr}$, where the moment aligns back to its CEF easy direction are also listed.  We also report their in-plane Ce-Ce distance ($d^\perp_{Ce\mbox{-}Ce}$, distance of Ce atoms within the same $ab$ plane), out-of-plane Ce-Ce distances ($d^\parallel_{Ce\mbox{-}Ce}$, distance of Ce atoms with a different $ab$ plane, often diagonally),  and their differences as $|\Delta d_{Ce\mbox{-}Ce}|$. For the compounds with orthorhombic symmetry or more than one Ce site, the values listed are the average of the nearest-neighbor distances. 

\begin{figure}[!htb]
\center
\includegraphics[width=\linewidth]{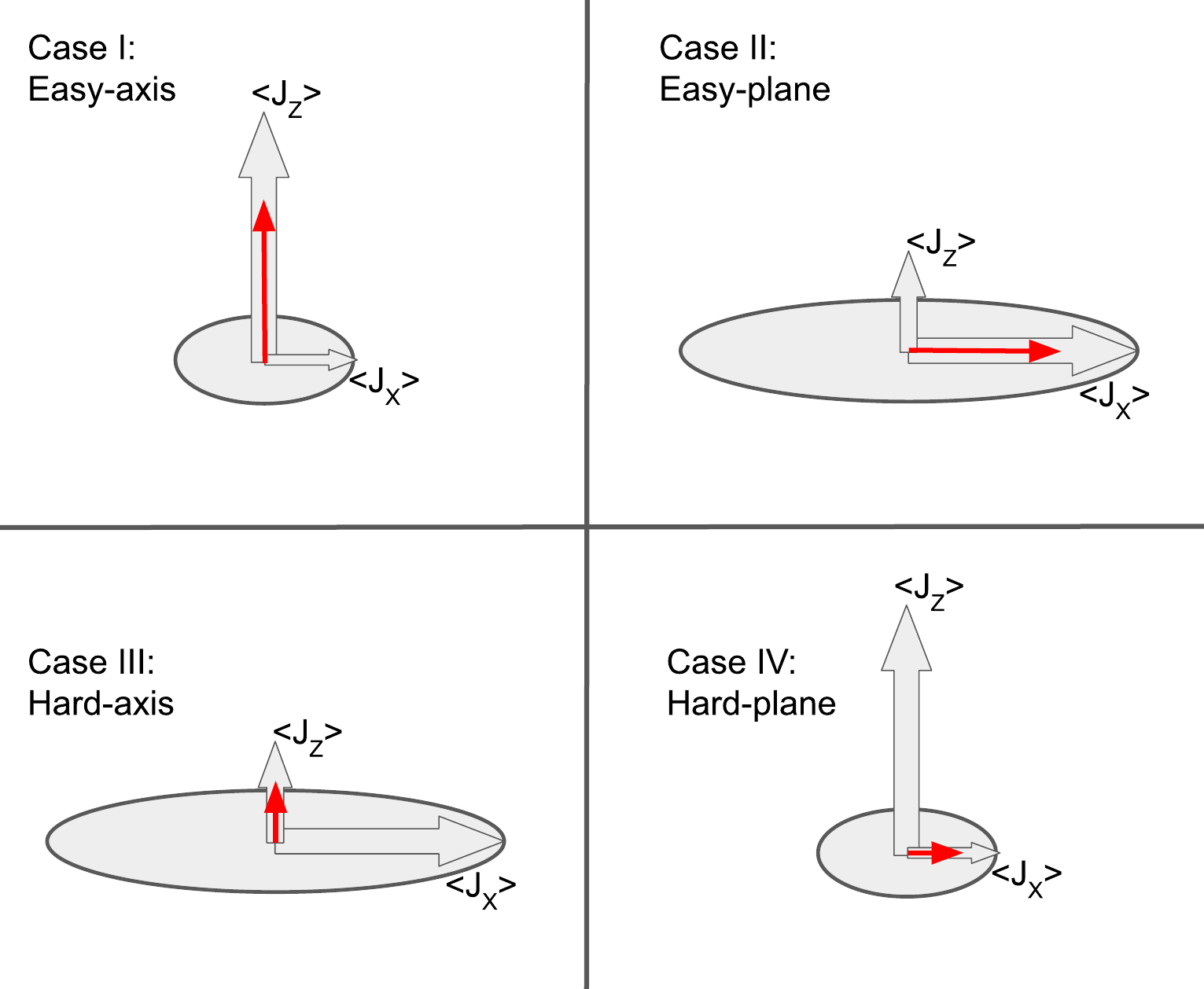}
\caption[]{Illustration of the four different categories based on the CEF ground-state moment (gray arrows) and the direction of the moment ordering (red arrow) at low temperature and low field.}
\label{CeFM_scheme}
\end{figure}

When a CEF scheme has already been reported, or when the CEF ground state is known, we also calculate the CEF anisotropy as $\Braket{J_x}$/$\Braket{J_z}$ at the ground state~\cite{hafner2019Kondolattice}. With this definition, the CEF anisotropy $>$1 favors the $ab$ plane, while $<$1 favors the $c$ axis. We also listed the molecular field contribution $\lambda_{i}$ if the data is available. For the compounds that have the available temperature-dependent inverse magnetization data, but without a detailed CEF analysis, we performed the CEF fitting. A detailed discussion of those compounds and their analysis can be found in the Appendix. The survey results are shown in Table~\ref{CeFMKLs}, and we group them into four different categories based on the direction of their magnetic ordering. Easy-axis or easy-plane ordering refers to magnetic ordering along the CEF ground state easy axis or plane, whereas hard-axis or hard-plane ordering occurs when the magnetic moments align along the CEF ground-state hard-axis or plane. The illustration of the four different categories is shown in Fig.~\ref{CeFM_scheme}.

Our survey shows that 25 Ce KLs exhibit FM ordering along the CEF easy axis or easy plane, while 5 are found to order along the CEF hard axis or hard plane. This suggests that hard-axis ordering is a rare occurrence in Ce systems, although it has been observed in several Ce or Yb compounds~\cite{hafner2019Kondolattice}. Based on the data aggregated in Table~\ref{CeFMKLs}, we propose that the occurrence of hard-axis or hard-plane ordering in Ce-based FM KLs primarily arises from the AFM interaction along the CEF easy axis or easy plane.

\begin{figure}[!htb]
\center
\includegraphics[width=\linewidth]{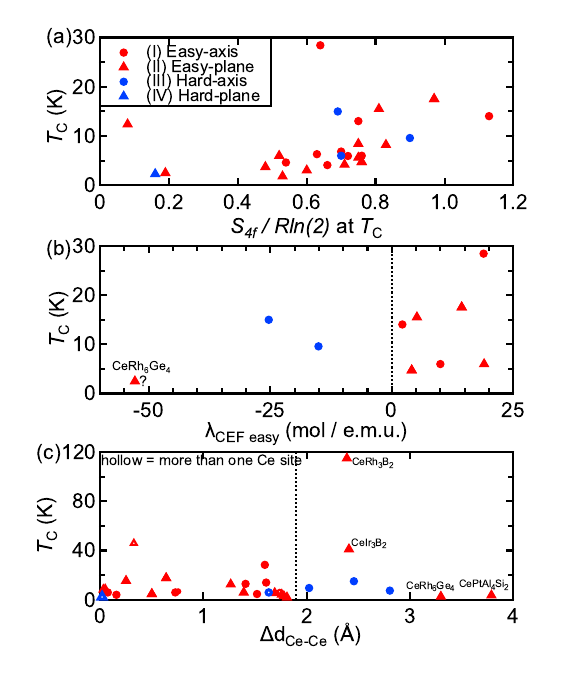}
\caption[]{Distribution of the listed Ce-compounds in Table~\ref{CeFM_tetra_table} based on (a) $S_{4f}/R\ln2$ at $T_\textrm{C}$. (b) Distribution based on the molecular field contribution at their CEF easy axis or easy plane ($\lambda_{\textrm{CEF easy}}$). Ce$_3$Al$_{11}$ and CeRu$_2$Al$_2$B are not shown because of their AFM ordering. (c) Distribution based Ce-Ce in-plane-out-of-plane distance difference $\Delta d_{\textrm{Ce}\mbox{-}\textrm{Ce}} = | d^\perp_{\textrm{Ce}\mbox{-}\textrm{Ce}} - d^\parallel_{\textrm{Ce}\mbox{-}\textrm{Ce}}|$.}
\label{Table_into_Fig}
\end{figure}

The entropy of Ce-based FM KLs at $T_\textrm{C}$ is expected to be $R\ln2$ because of the ground state doublet. However, the Kondo screening often reduces the entropy below this value. Thus, the ratio $S_{4f}/R\ln2$ at $T_\textrm{C}$ serves as a useful metric for estimating the strength of the Kondo effect. A lower value represents a strong Kondo screening effect. Figure~\ref{Table_into_Fig}(a) presents the visual distribution of the listed Ce-based FM KLs based on their $S_{4f}/R\ln2$ at $T_\textrm{C}$. We do not observe a clear correlation between the strength of the Kondo effect and whether a compound adopts hard-axis or hard-plane ordering.

Figure~\ref{Table_into_Fig}(b) presents the distribution based on their molecular field contribution along the CEF easy axis or easy plane, $\lambda_\textrm{CEF easy}$, for cases where reliable CEF analysis is available. All easy-axis or easy-plane ordering compounds exhibit a positive $\lambda_\textrm{CEF easy}$, with the exception of CeRh$_6$Ge$_4$. In contrast, the hard-axis ordering compounds, CeAgSb$_2$ and CeRuPO, show negative $\lambda_\textrm{CEF easy}$ values, indicating significant AFM interactions along the CEF easy-plane.

These results suggest that the hard-axis ordering observed in these two compounds arises from a combination of AFM interactions along the CEF easy-plane and FM interactions along the CEF hard-axis. This interpretation is consistent with Monte Carlo simulations based on the Heisenberg model for Yb(Rh$_{0.73}$Co$_{0.27}$)$_2$Si$_2$, which incorporate competing anisotropies in both exchange interactions and $g$ factors. These simulations successfully reproduce the experimental results and emphasize the importance of in-plane antiferromagnetic correlations~\cite{andrade2014Competing}.

CeRh$_6$Ge$_4$ is the only compound that does not follow the general trend. Although it exhibits easy-plane ordering, it has a negative $\lambda_{ab}$~\cite{shu2021Magnetic}. However, the negative $\lambda_i$ may not necessarily indicate that the RKKY exchange interactions are AFM. The smaller magnetization (only 0.34$\mu_B/$Ce at 5\,T, 1.8\,K) and the exceptionally low $S_{4f}$ at $T_\textrm{C}$ (only 0.19$R\ln2$) suggest a strong Kondo screening of the magnetic moments~\cite{matsuoka2015Ferromagnetic}. This might explain why both molecular field contributions $\lambda_i$ are negative, despite the compound being ferromagnetic. It is also possible that alternative molecular field contributions $\lambda_i$ could be obtained if the fitting was constrained with other measurements (see discussion in Appendix~\ref{CeRh6Ge4_appendix}).

Our survey suggests that the hard-axis ordering phenomena is mainly due to the anisotropic RKKY interactions, where the sign of the RKKY interaction along the CEF easy-axis and hard-axis are different. 
Figure~\ref{Table_into_Fig}(c) shows the distribution of compounds based on the difference between in-plane and out-of-plane Ce–Ce distances, defined as $\Delta d_{Ce\mbox{-}Ce} = | d^\perp_{Ce\mbox{-}Ce} - d^\parallel_{Ce\mbox{-}Ce}|$. We observe that compounds exhibiting easy-axis or easy-plane ordering generally have small differences in the in-plane versus the out-of-plane Ce-Ce distances (typically $<$ 1.9\,\AA), with a few exceptions (CeRh$_6$Ge$_4$, CePtAl$_4$Si$_2$, CeRh$_3$B$_2$, and CeIr$_3$B$_2$). Conversely, compounds demonstrating hard-axis ordering either have larger differences ($>$ 1.9\,\AA) in Ce-Ce distances (CeAgSb$_2$, CeRuPO, CeFeAs$_{0.7}$P$_{0.3}$O), or have more than one Ce site (Ce$_{2}$PdGe$_{3}$, $\beta$-CeNiSb$_3$). This large difference in the Ce-Ce distance may contribute to RKKY interactions with opposite signs along different crystallographic directions. 

It is not surprising that the anisotropic RKKY interaction naturally arises from the anisotropies of the Fermi surface and the anisotropies of the Ce-Ce distances since the anisotropic RKKY interaction is mediated by the conduction electrons, which reside at the Fermi surface of the material. Notably, even in cubic CeB$_6$, highly anisotropic RKKY interactions have been reported, originating from the multipole interactions~\cite{hanzawa2019Origin,kubo2004Octupole}. Such anisotropic RKKY interactions, arising from the Fermi surface anisotropy, have also been observed at the atomic scale in Co atoms~\cite{weismann2009Seeing, zhou2010Strength}.

At the same time, we cannot rule out the possibility of an anisotropic Kondo effect, which has been observed in other systems~\cite{fujii1989Anisotropic, otte2008Role, kanasz-nagy2018Exploring}. The significant AFM interaction along the CEF easy plane in hard-axis ordering compounds could be partially attributed to the anisotropic Kondo effect as well.

Therefore, based on our analysis, we believe that the hard-axis or hard-plane orderings in Ce-based FM KLs are primarily due to the AFM interaction along the CEF easy axis or easy plane. Our findings also suggest that an anisotropic RKKY model should be considered when analyzing Ce-based or, more broadly, heavy fermion systems to better understand their complex magnetic properties.

%%%%%%%%%%%%%%%%%%%%%%%%%%%%%%%%

%%%%%%%%%%%%%%%%

%%%%%%%%%%%%%%%%%%%%%%%%%%%%%%%
%%%%%%%%%%%%%%%%%%%%%%%%%%%%%%%%%%%%%%%%%%%%%%%%%%%%%%%%%%%%%%%%%%%%%%%%%%%%%

%%%%%%%%%%%%%%%%%%%%%%%%%%%%%%%%%%%%%%%%%%%%%%
\section{Further Discussion}

\subsection{\texorpdfstring{$\textbf{CeCuSi}$ versus $\textbf{CeAgSb}_2$}{CeCuSi vs CeAgSb2} }
Hexagonal CeCuSi crystallizes in a ZrBeSi-type structure with trigonal Ce site symmetry, and it orders ferromagnetically below $T_\textrm{C} = 15.5$\,K. A recent study on single-crystal CeCuSi indicates that its magnetic moments favor the CEF easy-plane, and the CEF model can well explain its magnetic properties~\cite{jin2024Easyplane}. Tetragonal CeAgSb$_2$ is also another example of a FM KL material, with $T_\textrm{C} = 9.6$\,K. CeAgSb$_2$ has attracted a lot of attention because of its remarkable magnetic hard-axis ordering~\cite{myers1999Systematic,araki2003Crystal,jobiliong2005Magnetization,nikitin2021Magnetic}. Although its CEF ground state favors the CEF easy-plane, the moments order along the hard-axis at low applied magnetic field. 

The physical properties of CeCuSi and CeAgSb$_2$ are listed in Table~\ref{PPcompare}, and these two compounds have many similarities. Both compounds exhibit FM ordering with a similar magnon gap energy determined from resistivity measurements, indicating the compounds are highly anisotropic. Both have a strong CEF effect with CEF ground states favoring the easy-plane, and the saturation moments can be well explained by the CEF ground states. Both exhibit a relatively weak Kondo effect based on the entropy value at $T_\textrm{C}$. The Kondo temperature $T_\textrm{K}$ can be estimated from the magnetic entropy $S_m$ with $S_m(0.5T_\textrm{K}) = 0.5R\ln2$~\cite{hafner2019Kondolattice}, and the ratio of $T_\textrm{K}$/$T_\textrm{C}$ is about 1.66 for both compounds. The $S_{4f}$ values at $T_\textrm{C}$ suggest that the Kondo effect is slightly stronger in CeCuSi than CeAgSb$_2$. This comparison further illustrates that there is no clear correlation between the Kondo effect and the hard-axis phenomenon.

%The molecular field contributions $\lambda_i$ were obtained based on Eq.~\ref{eq:inverchi}. The CEF anisotropy is defined as $\Braket{J_x}$/$\Braket{J_z}$ at the ground state~\cite{hafner2019Kondolattice}. 
The detailed CEF analysis for CeCuSi and CeAgSb$_2$ can be found in the previous study~\cite{jin2024Easyplane} and Appendix~\ref{CeAgSb2}, respectively. Our analysis suggests that CeAgSb$_2$ has a significant AFM interaction along the CEF easy-plane that potentially prevents the moments from aligning along the $ab$ plane, while there is no such AFM interaction for CeCuSi. 

\begin{table}[!htb]
\caption[Comparison of some of the physical properties of CeCuSi and CeAgSb$_2$]{Comparison of some of the physical properties of CeCuSi~\cite{jin2024Easyplane} and CeAgSb$_2$. $T_\textrm{K}$ is estimated based on $S_m(0.5T_\textrm{K}) = 0.5R\ln2$. CEF anisotropy = $\Braket{J_x}$/$\Braket{J_z}$.}
\renewcommand{\arraystretch}{1.2}
\begin{tabular}{p{0.33\linewidth}|p{0.26\linewidth}p{0.05\linewidth}p{0.21\linewidth}p{0.06\linewidth}}
              & CeCuSi &  & CeAgSb$_{2}$ &   \\ \hline
Space group         & $P6_{3}/mmc$ &     & $P4/nmm$  & \cite{brylak1995Ternary} \\
T$_C$               & 15.5\,K &     & 9.6\,K & \cite{myers1999Systematic} \\

$\theta _{CW}^{\parallel}$    & -52.6\,K &     &  -69.6\,K       & [\ref{CeAgSb2}] \\
$\theta _{CW}^{\perp}$    & 5.67\,K &     &  0.20\,K  & [\ref{CeAgSb2}] \\
$B_{2}^{0}$ from Eq.(\ref{B20CW})   & 4.35\,K &     &     6.51\,K & \\
$B_{2}^{0}$ from $\chi$ fitting   & 4.42\,K  &     &  7.55\,K  & \cite{takeuchi2003Anisotropic} \\
$B_{2}^{0}$ from INS   & 2.43\,K  & \cite{sondezi-mhlungu2009Crystal}~ & 7.6\,K & \cite{araki2003Crystal} \\
Ground state    &   $\sim 0.967\ket{\pm 1/2}$     &     &  $\ket{\pm 1/2}$    &  \\
CEF gaps    &   $\Delta_1 = 112$\,K     &     &  $\Delta_1 = 60.3$\,K    &  \\
            & $\Delta_2 = 122$\,K &     & $\Delta_2 = 145$\,K & \cite{araki2003Crystal}\\
& & & & \\
Magnon gap in $\rho$ (Eq.(4) in Ref.~\cite{jin2024Easyplane})   &  $\Delta = 23.9$\,K  &     & $\Delta = 24.4$\,K  & \cite{jobiliong2005Magnetization} \\
$A$\,($\mu\Omega\,\mathrm{cm}\,\mathrm{K}^{-2}$)   &   0.019     &     & 0.08  & \cite{jobiliong2005Magnetization} \\
$B$\,($\mu\Omega\,\mathrm{cm}\,\mathrm{K}^{-2}$)   &   0.133     &     & 3.3  & \cite{jobiliong2005Magnetization} \\
& & & & \\
$\gamma_{4f}$\,(mJ mol$^{-1}$K$^{-2}$)       &   41.2     &     & 46  & \cite{takeuchi2003Anisotropic} \\
$S_{4f}$ at $T_\textrm{C}$    &  $0.81 R \ln 2$  &     & $0.9 R \ln 2$    &  \cite{takeuchi2003Anisotropic} \\
Estimated $T_\textrm{K}$    &  25.6\,K  &     & 16\,K    &  \cite{hafner2019Kondolattice,takeuchi2003Anisotropic} \\
$\lambda^{\parallel}$ \,(mol/e.m.u.)    &  9.0  &     & 12.07   & [\ref{CeAgSb2}]\\
$\lambda^{\perp}$ \,(mol/e.m.u.)    &  5.2  &     & -15.06    &[\ref{CeAgSb2}]\\
CEF anisotropy   &  4.6  &     & 3   &\cite{hafner2019Kondolattice}
\end{tabular}
\label{PPcompare}
\end{table}

\subsection{Hard-plane ordering: \texorpdfstring{Ce$_{2}$PdGe$_{3}$}{Ce2PdGe3}}
Ce$_{2}$PdGe$_{3}$~\cite{baumbach2015Complex, bhattacharyya2016Exploring} is the only compound in our survey that exhibits hard-plane ordering. The high-temperature susceptibility indicates that the CEF easy axis is along the $c$ axis, while the low-temperature magnetization below $T_\textrm{C} = 2.25$\,K reveals that magnetic ordering occurs in the $ab$ plane. This compound also exhibits AFM transitions at 10.7\,K and 9.6\,K, observed only along the $c$ axis. These transitions suppress the magnetic moments along the $c$-axis below 10.7\,K, ultimately leading to hard-plane ordering~\cite{baumbach2015Complex}. A neutron diffraction study further suggests that the AFM ordering around 11\,K is along the $c$ axis, and the FM ordering is within the tetragonal basal plane around 2.5\,K~\cite{bhattacharyya2016Exploring}. 

This compound highlights the importance of both FM and AFM interactions. Anisotropic RKKY interactions must be considered to understand their complex magnetic behavior. We did not perform CEF analysis for this compound due to the presence of two inequivalent Ce sites.

%%%%%%%%%%%%%%%%%%%%%%%%%%%%%%%%%%%%%%%%%%%%%%%%%%%%%%%%%%%%%%%%%%%%%%%%%%%%%
\subsection{Hard-plane AFM ordering: CeAlGe}

\begin{figure*}[!htb]
\center
\includegraphics[width=\linewidth]{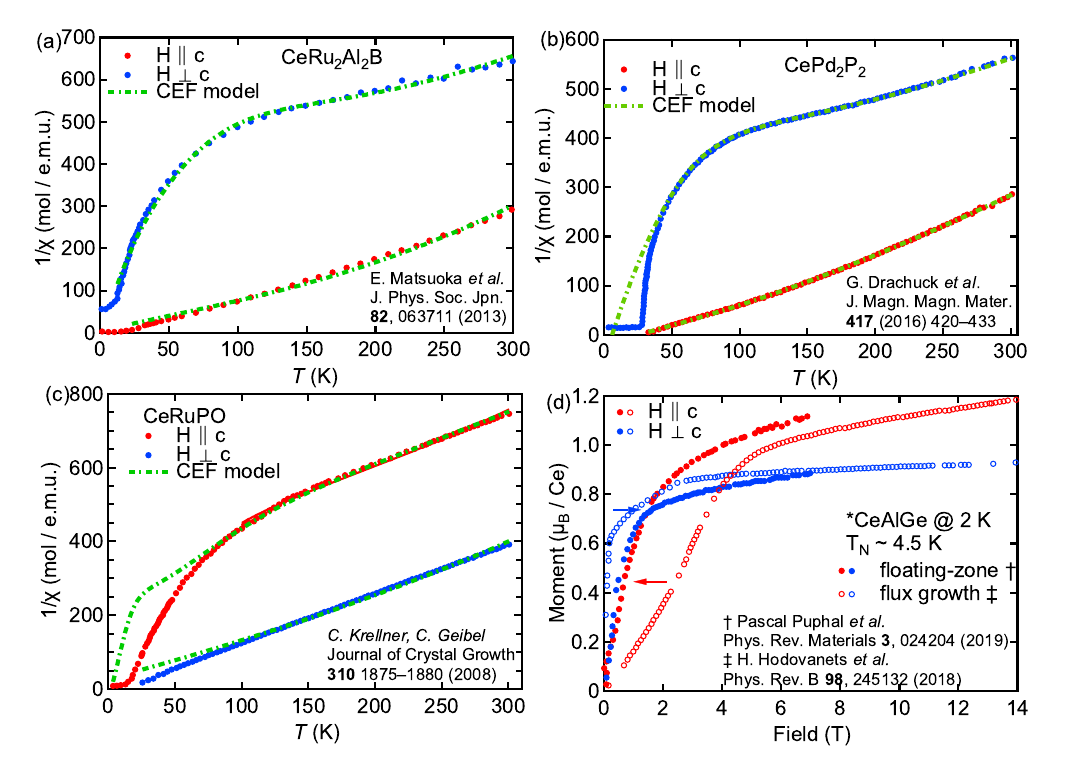}
\caption[]{Temperature-dependence of inverse magnetic susceptibility along the $c$ axis and $ab$ plane of tetragonal compounds:(a) CeRu$_2$Al$_2$B[\ref{CeRu2Al2B_appdix}], (b) CePd$_2$P$_2$[\ref{CePd2P2_appdix}], and (c) CeRuPO[\ref{CeRuPO_appdix}]. The green dash lines are the fitting based on the CEF model and fitting parameters can be found in Table~\ref{CeFM_tetra_table}. (d) Field-dependent magnetization of AFM ordering compound CeAlGe synthesized using two different methods.}
\label{CeFM_tetra_figs}
\end{figure*}

\begin{table*}[!hpt]
\caption{Fitting parameters for the inverse magnetic susceptibility data on tetragonal Ce-based compounds based on Eqs.~(\ref{eq:chipara_tetra}) and (\ref{eq:chiperpen_tetra}).}

\begin{tabular}{l|l|l|l|p{0.10\linewidth}|p{0.10\linewidth}|p{0.10\linewidth}|p{0.10\linewidth}|l}
 & $B_{2}^{0}$ (K) & $B_{4}^{0}$ (K)   & $B_{4}^{4}$ (K)  & $\lambda_{\parallel}$ (mol/e.m.u) & $\lambda_{\perp}$ (mol/e.m.u) & $\chi_{0}^{\parallel} \times 10^{-5}$ (e.m.u./mol)          & $\chi_{0}^{\perp} \times 10^{-5}$ (e.m.u./mol)& Case\\ \hline

CeRu$_2$Al$_2$B~\cite{matsuoka2013Magnetic},[\ref{CeRu2Al2B_appdix}] & -30   &    -1   &  -8  &  3.06 $\pm$ 3.3     &    -36.9 $\pm$ 0.91    &   -88.9 $\pm$ 8.7    &  19.3 $\pm$ 2.2 & (I)\\ 
CePd$_2$P$_2$~\cite{drachuck2016Magnetization},[\ref{CePd2P2_appdix}]& -18.8 $\pm$ 0.12 & -2.23$\pm$0.08 & 9.47$\pm$0.16 & 18.9$\pm$0.2      &      59.5 $\pm$ 4.3          & -28.6 $\pm$ 0.6 &  -8.2 $\pm$ 0.6 &  (I)   \\
CeRuPO~\cite{krellner2008Single},[\ref{CeRuPO_appdix}] &  27.8$\pm$1.26  &  0.24$\pm$0.02  &   3.79$\pm$0.69   &   33.1 $\pm$ 7.46    &        -25.3 $\pm$ 1.90 &  -21 $\pm$ 3.3   &  -49 $\pm$ 2.8&  (III)\\
\end{tabular}
\label{CeFM_tetra_table}
\end{table*}

During our survey, we identified a compound with AFM hard-plane ordering, CeAlGe, which offers further insights. Despite exhibiting net magnetization in the $ab$ plane at low fields, CeAlGe is determined to have an AFM order, as evidenced by the absence of an anomalous Hall effect~\cite{hodovanets2018Singlecrystal}. The unique features in its ac susceptibility measurements are consistent with
a spin-flop transition~\cite{hodovanets2018Singlecrystal,balanda2013AC}. Recent neutron diffraction results also support the in-plane ferromagnetic component at low field~\cite{suzuki2019Singular}. This compound is another example that highlights the presence of anisotropic RKKY interactions.

Single crystals of CeAlGe can be synthesized using either the flux growth method~\cite{hodovanets2018Singlecrystal} or the floating-zone method~\cite{puphal2019Bulk}. However, their exact stoichiometries differ, and it is known that the physical properties of this system are sensitive to precise stoichiometry~\cite{dhar1996Structural}. EDS analysis indicates that the floating-zone method yields a composition closer to the target ratio, while the flux method results in Ge-deficient CeAl$_{1.12}$Ge$_{0.88}$~\cite{puphal2019Bulk}. Field-dependent magnetization data from both methods are shown in Fig.~\ref{CeFM_tetra_figs}(d). As the stoichiometry ratio changes from Ge-deficient to the balanced ratio, the magnetization curves along the $ab$ plane and the $c$ axis become more isotropic, and the crossing field decreases from nearly 4\,T to 1.5\,T. If this trend continues, a Ge-rich version of CeAlGe may no longer exhibit hard-axis ordering behavior. 

A similar synthesis-method-dependent physical properties difference in this compound was reported in Ref.~\cite{hodovanets2022Anomalous}. The level of Al deficiency in the samples also affects the crossing field. 
The Al flux-grown sample (using cerium ingot from Ames Laboratory) has more Al deficiency with a composition of CeAl$_{0.83}$Ge$_{0.96}$, and a crossing field about 4.4\,T. The In flux-grown sample (using cerium ingot from Alfa Aesar) has less Al deficiency with a composition of CeAl$_{0.90}$Ge$_{0.98}$, and a crossing field of about 3.3\,T~\cite{hodovanets2022Anomalous}.

The CEF effects in two different stoichiometries of CeAlGe are expected to be similar because they maintain the same crystal structure with minimal lattice parameter variations and slight changes in Al/Ge stoichiometry~\cite{puphal2019Bulk}. The Kondo effect is likewise comparable in both stoichiometries because the stoichiometry of the Ce element remains constant, and the resistivity measurements for both stoichiometries are also very similar. Therefore, the dramatic change in magnetization is likely due to the exchange interaction itself, supporting the argument that hard-axis ordering is primarily due to anisotropic exchange interaction.

In CeAlGe, the ferromagnetic component at low field enables the evidence of hard-plane ordering with a crossing in magnetization-versus-field curves. However, a magnetization crossing is not always expected for a pure AFM ordering because AFM ordering reduces the magnetic susceptibility. This makes the identification of hard-axis or hard-plane AFM ordering more difficult. Antiferromagnetic compounds are beyond the scope of this study, but hard-axis ordering has been observed in antiferromagnetic compounds as well. In CeOs$_2$Al$_{10}$ and CeRu$_2$Al$_{10}$,the ordered moment is along the $c$ axis~\cite{khalyavin2010Longrange,kato2011Magnetic} while the CEF ground state favors the $a$ axis~\cite{strigari2013Crystal,adroja2013Muonspinrelaxation}.

%%%%%%%%%%%%%%%%%%%%%%%%%%%%%%%%%%%%%%%%%%%%%%%%%%%%%%%%%%%%%%%%%%%%%%%%%%%%%
%%%%%%%%%%%%%%%%%%%%%%%%%%%%%%%%%%%%%%%%%%%%%%%%%%%%%%%%%%%%%%%%%%%%%%%%%%%%%

% Place all of the references you used to write this paper in a file
% with the same name as following the \bibliography command
%\bibliography{CeCuSi}

%%%%%
\section{Conclusion}
In summary, we conducted a detailed survey of Ce-based FM-KL compounds with available single-crystal magnetization data. Our survey suggests that hard-axis or hard-plane ordering is relatively uncommon. All Ce-based FM KLs with observed hard-axis or hard-plane ordering either exhibit a large difference between in-plane and out-of-plane Ce–Ce distances($>$ 1.9\,\AA) or have more than one inequivalent Ce site. We did not find a clear correlation between the strength of the Kondo effect and the occurrence of hard-axis or hard-plane ordering.

The results of our survey indicate that the interplay between CEF anisotropy and anisotropic exchange interactions is the primary factor determining the magnetic ordering direction. The comparison between CeCuSi and CeAgSb$_2$ supports our conclusion, suggesting that hard-axis ordering in Ce-based FM KLs is primarily driven by the AFM interaction along the CEF easy-plane. Our findings also highlight the importance of considering anisotropic RKKY models when analyzing the complex magnetic properties of Ce-based compounds and other heavy fermion systems.

%%%%%%%%%

\section{Acknowledgement}
We thank Anton Jesche and Manuel Brando for useful discussions. We acknowledge support from the Department of Physics and Astronomy, University of California, Davis, U.S.A. and from the Cahill research fund. We also acknowledge support from the Physics Liquid Helium Laboratory fund.

\section{Data Availability}

The data that support the findings of this paper are openly available~\cite{Jin2025PRBdataset}.

%%%%

\appendix
%%%%%%%%%%%%%%%%%%%%%%

\section{Limitations of the CEF analysis of paramagnetic susceptibility}\label{limitation}

(1) There are often multiple sets of CEF parameters ($B_m^n$) that can fit high-temperature magnetization data equally well. To accurately determine the CEF parameters, it is important to apply additional constraints that help narrow down the possible solutions. Such constraints include CEF level splitting energies obtained from Schottky anomalies, inelastic neutron scattering, soft x-ray absorption, and the ground state inferred from saturation moments in field-dependent magnetization measurements. Without these constraints, it becomes very difficult to reliably determine CEF parameters solely from fitting inverse magnetic susceptibility data. This problem worsens when the additional parameters of the mean-field interactions ($\lambda_i$) are considered, because they act on the paramagnetic susceptibility similarly to the $B_2^0$ coefficient.

Magnetic anisotropy mainly consists of two contributions: CEF anisotropy and exchange anisotropy.  For simplification, if we neglect the CEF anisotropy from $B_{m}^{n}$ terms other than $B_{2}^{0}$, then the magnetic anisotropy reflected in the Curie-Weiss temperature can be interpreted using the following expression~\cite{boutron1973Exact}:
\begin{equation} \label{exchangeJ}
\begin{split}
\theta _{\textrm{CW}}^{\parallel}&=-\frac{2J(J+1)}{3k_{B}}J_{\textrm{ex}}^{\parallel}-\frac{(2J-1)(2J+3)}{5k_{B}}B_{2}^{0},\\
\theta _{\textrm{CW}}^{\perp}&=-\frac{2J(J+1)}{3k_{B}}J_{\textrm{ex}}^{\perp}+\frac{(2J-1)(2J+3)}{10k_{B}}B_{2}^{0}.
\end{split}
\end{equation}
In this approximation, $\lambda_{i} \approx \frac{2J(J+1)}{3k_{B} \cdot \mathcal{C}}J_{\textrm{ex}}^{i}$, where $\mathcal{C}$ is the Curie constant, assuming the contributions from higher-order $B_{m}^{n}$ terms are small. For the Ce$^{3+}$ system, $J=5/2$ and $g_J = 6/7$, which gives a Curie constant $\mathcal{C} = 0.804$\,(e.m.u.$\cdot$K/mol). 

Using the above equations and approximation, a change in $B_{2}^{0}$ leads to a change in anisotropy of the Curie-Weiss temperature : $\theta_{ab}-\theta_{c}=9.6B_{2}^{0}$ with $B_{2}^{0}$ in K. But similarly, a change in molecular field contribution  leads to : $\theta_{ab}-\theta_{c}=0.804 (\lambda_{\perp}-\lambda_{\parallel})$ with $\lambda_i$ in mol/e.m.u. As a result, $B_{2}^{0}$ and $\lambda_i$ cannot be determined independently when only using the high temperature susceptibility: every 1\,K change in $B_{2}^{0}$ is equivalent to a difference of 11.9\,(mol/e.m.u.) in $|\lambda_{\perp}-\lambda_{\parallel}|$.

%Thus, obtaining reliable $\lambda_i$ values from CEF analysis relies on having a consistent CEF scheme that can coherently explain all relevant physical properties, including CEF energy splitting and ground state behavior.

We illustrate the situation with the case of CeAgSb$_2$. In Fig.~\ref{CeAgSb2_fig}, we show three models that can fit the magnetic susceptibility of CeAgSb$_2$ relatively well, but with significantly different values of $B_{2}^{0}$. The CEF parameters for the three models as well as from previous studies~\cite{araki2003Crystal,jobiliong2005Magnetization} are listed in Table~\ref{CeAgSb2_table}. In model 2, we fixed $B_{2}^{0}$ at 1\,K and obtained the remaining parameters through fitting. Although the fit looks good, this model predicts opposite signs for $\lambda_i$ and a ground state of $\Gamma_{7}^{(1)}$ instead of $\Gamma_6$, leading to a very different interpretation of the magnetic anisotropy. In model 3, we kept $B_{4}^{0}$ and $B_{4}^{4}$ the same as in model 2 but fixed $B_{2}^{0}$ at 10\,K. This also resulted in a decent fit, but also a $\Gamma_{7}^{(1)}$ ground state, along with significantly different CEF splitting energies. However, both inelastic neutron scattering and the Schottky anomaly suggest that the CEF splitting energies should be approximately  $\Delta_1 \sim 60$\,K and $\Delta_2 \sim 140$\,K~\cite{araki2003Crystal,takeuchi2003Anisotropic}. In addition, the saturation moment observed in higher-field measurements supports a $\Gamma_6$ ground state~\cite{takeuchi2003Anisotropic,jobiliong2005Magnetization}. Therefore, despite their reasonable fits to inverse magnetic susceptibility, CEF models 2 and 3 are inconsistent with other key experimental observations. More details about CEF analysis in CeAgSb$_2$ can be found in Appendix~\ref{CeAgSb2}. This example illustrates that CEF analysis based solely on inverse magnetic susceptibility may not be sufficient to distinguish the magnetic anisotropy contributions from $B_{2}^{0}$ and $\lambda_i$. A reliable CEF model should consistently account for all physical properties arising from CEF energy level splitting and correctly identify the CEF ground state.

(2) A good CEF analysis on single crystal magnetization data requires a well-centered and well-mounted sample. When measuring FM compounds using the Quantum Design MPMS, the sample must be positioned at the center (radially) of the straw to obtain an accurate signal. If the sample is not well-centered, it may lead to a ``false'' hard-axis ordering~\cite{ullah2024Experimental}.

(3) During the fitting process, the role of residual susceptibility $\chi_0$ can be significant when the magnetic susceptibility value is small. This typically occurs with the CEF hard axis or hard plane, potentially leading to inaccurate evaluations of $\lambda_{\textrm{hard}}$. Therefore, it is essential to either use a consistent glue/grease-free mounting method or perform background subtraction to reduce background susceptibility contributions.

(4) When choosing a different set of CEF parameters, the variation in $\lambda_{\textrm{easy}}$ is much smaller than that in $\lambda_{\textrm{hard}}$. The magnetic susceptibility in the hard axis or hard plane is smaller, i.e., more sensitive to the change of the CEF parameters and $\chi_0$ values. This means that the error on $\lambda_{\textrm{hard}}$ is higher than that on $\lambda_{\textrm{easy}}$.

(5) The CEF model assumes a localized point-charge framework and does not account for effects such as Kondo screening, hybridization of localized moments, or magnetic ordering, which are common in heavy fermion systems. As a result, this model may not perfectly reproduce the magnetization data, particularly in the low-temperature region.

\section{\texorpdfstring{Tetragonal CEF in CeAgSb$_2$ }{Tetragonal CEF in CeAgSb2}} \label{CeAgSb2}

Samples of CeAgSb$_2$ were grown using the flux growth method~\cite{myers1999Systematic}. A sample was cut into a bar shape to ensure a uniform mounting method along the $c$ axis and the $ab$ plane in the MPMS, without the use of grease or glue, to minimize the residual susceptibility $\chi_0$. Additionally, the sample was carefully polished to remove any residual flux from the surface.

\begin{figure}[!t]
\center
\includegraphics[width=\linewidth]{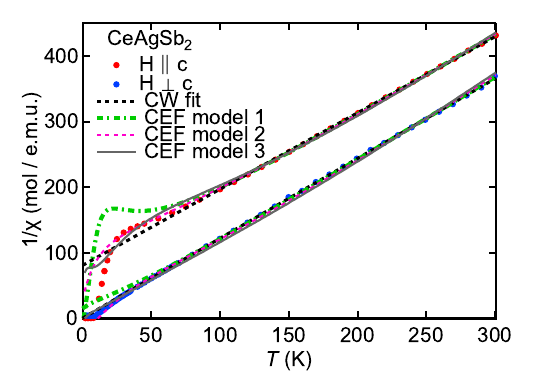}
\caption{Temperature-dependence of inverse magnetic susceptibility of CeAgSb$_2$ along the $c$ axis and $ab$ plane with $\mu_0H = 0.1$\,T. The black and green dashed lines are fits to the Curie-Weiss law and the CEF model 1 (see Appendix~\ref{CeAgSb2}). CEF models 2 and 3 also provide good fits but use significantly different values of $B_{2}^{0}$ (see discussion in Appendix~\ref{limitation}).}
\label{CeAgSb2_fig}
\end{figure}

The inverse susceptibility as a function of temperature with an applied field of 0.1\,T along the $c$ axis and the $ab$ plane from 2\,K to 300\,K is shown in Fig.~\ref{CeAgSb2_fig}. Fits to the Curie-Weiss law are shown as black dashed lines. The effective moments and Curie-Weiss temperatures are 2.62\,$\mu_B$ and $-69.6$\,K along the $c$ axis, and are 2.54\,$\mu_B$ and $-0.20$\,K along the $ab$ plane, respectively.

For a Ce$^{3+}$ ion in a tetragonal site symmetry, the CEF Hamiltonian can be written as, 
\begin{equation} \label{eq:H_tetra}
H_{\mathrm{CEF}}=B_{2}^{0}O_{2}^{0}+B_{4}^{0}O_{4}^{0}+B_{4}^{4}O_{4}^{4},
\end{equation}
where $B_{m}^{n}$ and $O_{m}^{n}$ are the CEF parameters and the Stevens operators, respectively~\cite{stevens1952Matrix,hutchings1964Pointcharge}. 

Using the similar CEF analysis approach as in previous studies~\cite{banda2018Crystalline, jin2024Easyplane}, the final theoretical susceptibilities along both axes in the tetragonal point symmetry can be expressed as:
\begin{equation} \label{eq:chipara_tetra}
\begin{split}
\chi_{B\parallel z} =\frac{N_{A}g_{J}^{2}\mu _\mathrm{{B}}^{2}\mu_{0}}{Z} \Bigl\{ \frac{\beta}{4}\Big[1 + (5\sin^2\alpha-3\cos^2\alpha)^2e^{-\beta \Delta_{1}}\\+(5\cos^2\alpha-3\sin^2\alpha)^2e^{-\beta \Delta_{2}}\Big]\\ + \Big[ 32\sin^2\alpha \cos^2\alpha \cdot \frac{e^{-\beta \Delta_{1}}-e^{-\beta \Delta_{2}}}{\Delta_{2}-\Delta_{1}} \Big] \Bigr\},
\end{split}
\end{equation}

\begin{equation} \label{eq:chiperpen_tetra}
\begin{split}
\chi_{B\perp  z} =\frac{N_{A}g_{J}^{2}\mu _\mathrm{{B}}^{2}\mu_{0}}{Z} \Bigl\{ \frac{\beta}{4}\Big[9 + 20\sin^2\alpha\cos^2\alpha e^{-\beta \Delta_{1}}\\+20\sin^2\alpha\cos^2\alpha e^{-\beta \Delta_{2}}\Big]\\ + \frac{1}{2}\Big[8\cos^2\alpha \cdot
\frac{1-e^{-\beta \Delta_{1}}}{\Delta_{1}}+ 8\sin^2\alpha \cdot\frac{1-e^{-\beta \Delta_{2}}}{\Delta_{2}}\\ + 5(\sin^2\alpha-\cos^2\alpha)^2\cdot \frac{e^{-\beta \Delta_{1}}-e^{-\beta \Delta_{2}}}{\Delta_{2}-\Delta_{1}} \Big] \Bigr\},
\end{split}
\end{equation}
with 
\begin{equation}
\begin{aligned}
\Delta_{1} &= +12B_{2}^{0}-180B_{4}^{0} - S, \\
\Delta_{2} &= +12B_{2}^{0}-180B_{4}^{0} + S, \\
S &= \sqrt{(6B_{2}^{0} + 120B_{4}^{0})^2 +(12\sqrt{5}B_{4}^{4})^2},\\
\alpha &= \tan^{-1} \Big( \frac{6B_{2}^{0}+120B_{4}^{0}}{12\sqrt{5}\left|B_{4}^{4}\right|}-\sqrt{1+(\frac{6B_{2}^{0}+120B_{4}^{0}}{12\sqrt{5}\left|B_{4}^{4}\right|})^2} \Big),
\end{aligned}
\label{tetra_delta_relation}
\end{equation}

assuming the states will split into the following states from the ground state to the higher excited states:
\begin{equation}
\begin{aligned}
\Gamma_{6} &= \ket{\pm 1/2},\\
\Gamma_{7}^{(1)} &= \cos\alpha\ket{\pm 5/2} + \sin\alpha\ket{\mp 3/2}, \\
\Gamma_{7}^{(2)} &= \sin\alpha\ket{\pm 5/2} - \cos\alpha\ket{\mp 3/2}.
\end{aligned}
\end{equation}
Depending on the specific parameters, the ground state may vary, and the splitting $\Delta_i$ can be negative to indicate a different ground state.

\begin{table*}[!hpt]
\caption{Fitting parameters to the inverse magnetic susceptibility data based on Eqs.~(\ref{eq:chipara_tetra}) and (\ref{eq:chiperpen_tetra}) using previously reported CEF parameters for CeAgSb$_2$.}

\begin{tabular}{l|l|p{0.07\linewidth}|p{0.07\linewidth}|l|l|l|l|p{0.11\linewidth}|p{0.11\linewidth}}
 & $B_{2}^{0}$ (K) & $B_{4}^{0}$ (K)   & $B_{4}^{4}$ (K) & $\Delta_1$ (K)&$\Delta_2$ (K) & $\lambda_{\parallel}$ (mol/e.m.u.) & $\lambda_{\perp}$ (mol/e.m.u) & $\chi_{0}^{\parallel} \times 10^{-5}$ (e.m.u/mol)          & $\chi_{0}^{\perp} \times 10^{-5}$ (e.m.u/mol) \\ \hline

CEF model 1~\cite{takeuchi2003Anisotropic} &  7.55   &  -0.02  &   0.64   & 48.0 & 140.4 &   12.07 $\pm$ 0.81    &        -15.06 $\pm$ 0.61 &  -1.53 $\pm$ 0.68   &  -4.47 $\pm$ 1.09  \\
Ref.~\cite{araki2003Crystal} & 7.6    &    -0.06   &  0.7 & 59.3 & 144.7 &  16.82 $\pm$ 1.21     &    -15.23 $\pm$ 0.64    &   -3.49 $\pm$ 0.91    &  -4.67 $\pm$ 1.13  \\ 
Ref.~\cite{jobiliong2005Magnetization}& 6.6 & -0.09 & 1.14 & 53.4 & 137.4 & 4.75 $\pm$ 0.90      &      -12.49 $\pm$ 0.56          & -1.93 $\pm$ 0.77 &  -4.44 $\pm$ 1.05      \\

CEF model 2 [\ref{limitation}]& 1 & -0.1147 & 5.98 & -128.0\footnotemark[1] & 193.3 &   -27.45 $\pm$ 3.03    &      28.86 $\pm$ 1.49          & -5.51 $\pm$ 1.64 &  -12.27 $\pm$ 0.78      \\

CEF model 3 [\ref{limitation}]& 10 & -0.1147  & 5.98  & -26.4\footnotemark[1] & 307.7 &   84.87 $\pm$ 1.62    &      0.88 $\pm$ 0.80          & -14.53 $\pm$ 1.63 &  -17.50 $\pm$ 1.56      \\
\end{tabular}
\label{CeAgSb2_table}
\footnotetext[1]{Negative $\Delta_1$ means a ground state of $\Gamma_{7}^{(1)}$ instead of $\Gamma_{6}$}

\end{table*}

With the initial estimation of the $B_{2}^{0}$ from Eq.~(\ref{B20CW}), the experimental data were fitted to Eq.~(\ref{eq:inverchi}) with Eq.~(\ref{eq:chipara_tetra}) and Eq.~(\ref{eq:chiperpen_tetra}) for both orientations simultaneously. However, the fittings give a large range of $B_m^n$, and many of the corresponding CEF splitting energies do not match the results from neutron measurements. Therefore, we used several previously reported CEF parameters to fit our measurements, the results are listed in Table~\ref{CeAgSb2_table}. The magnitude of the residual susceptibility $\chi_0$ is only in the order of $\times 10^{-5}$ (e.m.u./mol), and the results would not be significantly affected if set to zero. The first set of parameters gives the best fit and is plotted as the green dashed lines (CEF model 1) in Fig.~\ref{CeAgSb2_fig}. We also plot two extreme cases (CEF models 2 and 3) with significantly different $B_{2}^{0}$ values to illustrate the limitations of our CEF fitting (as discussed previously in Appendix~\ref{limitation}).

In our analysis, regardless of the chosen parameters, the molecular field $\lambda_{\parallel}$ is always negative, and $\lambda_{\perp}$ is always positive, with the magnitudes smaller than 20\,mol/e.m.u. This result agrees with the discussion in the supplementary information of Ref.~\cite{hafner2019Kondolattice}. Although the inelastic neutron scattering (INS) results indicate that the interaction in the FM ordering region at low temperatures is overall ferromagnetic~\cite{araki2003Crystal, nikitin2021Magnetic}, our high-temperature susceptibility infers that there are antiferromagnetic interactions existing along the $ab$ plane. The origin of this AFM interaction could be simply from the RKKY interaction as we discussed in the earlier section. This interaction could prevent the magnetic moments from aligning along the CEF easy-plane at low temperature and low magnetic field.

%%%%%%%%%%%%%%%%

\section{\texorpdfstring{CeTiGe$_3$}{CeTiGe3}}
\label{CeTiGe3_appdix}

CeTiGe$_3$ is a rare FM KL with a pure ground state of $\ket{\pm 5/2}$, which leads to a CEF ground-state anisotropy value to be zero using our definition, indicating an Ising-like CEF ground-state anisotropy. Such a compound necessarily falls under case I. It is also the only compound where the $4f$ entropy contribution $S_{4f}$ at $T_\textrm{C}$ is larger than $R\ln2$~\cite{inamdar2014Anisotropic}. The CEF analysis can be found in Ref.~\cite{jin2022Suppression}.

\section{CeGaGe} \label{CeGaGe_appdix}

CeGaGe~\cite{ram2023Magnetic} is also a FM KL with $T_\textrm{C} = 6.0$\,K. This compound has a tetragonal structure with an easy-axis ordering. The earlier study on polycrystalline samples suggests that the structure has a centrosymmetric space group $I4_1/amd$ (space group 141), with a Ce point group of $D_{2d}$ (tetragonal)~\cite{grin1991Physical}. The recent PXRD study on the single-crystal samples suggests that the structure instead has a noncentrosymmetric space group $I4_1md$ (space group 109), with a Ce point group of $C_{2v}$(orthorhombic)~\cite{ram2023Magnetic}. A similar CEF analysis was performed, assuming a point symmetry of $C_{4v}$ (tetragonal) (without the $B_{2}^{2}$ and $B_{4}^{2}$ coefficients)~\cite{ram2023Magnetic}. The result shows that CeGaGe has a ground state of $0.92\ket{\pm 5/2}-0.38\ket{\mp 3/2}$, favoring an alignment along the $c$ axis. Likewise, the CEF model could effectively account for the curvature behavior observed in inverse magnetic susceptibilities, saturation moments, and Schottky anomalies in heat capacity. The CEF ground-state anisotropy is 0.42 (easy-axis), $\lambda_{\parallel} = 10.0$\,mol/e.m.u. and $\lambda_{\perp} = 18.7$\,mol/e.m.u.~\cite{ram2023Magnetic}.

%%%%%%%%%%%%%
%%%%%%%%%%
\section{\texorpdfstring{$\textbf{CeRu$_2$Al$_2$B}$ }{CeRu2Al2B}}\label{CeRu2Al2B_appdix}

The single crystal magnetization data for CeRu$_2$Al$_2$B were taken from Ref.~\cite{matsuoka2013Magnetic}, as shown in Fig.~\ref{CeFM_tetra_figs}(a). The previous studies suggest that the ground state of CeRu$_2$Al$_2$B is $\Gamma_7^{(1)}$~\cite{matsuno2012Isingtype}. If we use the reported CEF parameters of $B_{2}^{0} = -30$\,K, $B_{4}^{0} = -1$\,K, and $B_{4}^{4} = -8$\,K, then the refined fitting will give $\lambda_{\parallel} = 3.06$\,mol/e.m.u. and $\lambda_{\perp} = -36.9$\,mol/e.m.u. The fitting result is shown in Fig.~\ref{CeFM_tetra_figs}(a), and the parameters are listed in Table~\ref{CeFM_tetra_table}. This categorizes the compound as Case I.

%%%%%%%%%%%%%
%%%%%%%%%%
\section{\texorpdfstring{$\textbf{CePd$_2$P$_2$}$ }{CePd2P2}}\label{CePd2P2_appdix}

The single crystal magnetization data for CePd$_2$P$_2$ were taken from Ref.~\cite{drachuck2016Magnetization}, as shown in Fig.~\ref{CeFM_tetra_figs}(b). This compound also lacks an analysis of Schottky anomaly or INS data to narrow down the CEF parameters range. With an estimated $\Delta_{\mathrm{CEF}}=260\pm30$\,K from the previous resistivity measurements~\cite{tran2014Ferromagnetism}, the best fitting gives CEF parameters of $B_{2}^{0} = -18.8$\,K, $B_{4}^{0} = -2.23$\,K, and $B_{4}^{4} = -9.47$\,K. The fitting leads to a ground state of $\Gamma_7^{(1)}$ with a first CEF level at $\Delta_{\mathrm{CEF}}=282$\,K, and an expected $M^{\textrm{sat}}_{\parallel c} = 1.85$\,$\mu_B$. This is close to the observed $M^{\textrm{sat}}_{\parallel c} = 1.64$\,$\mu_B$~\cite{drachuck2016Magnetization}. The results are shown in Fig.~\ref{CeFM_tetra_figs}(b), and the parameters are listed in Table~\ref{CeFM_tetra_table}. The CEF ground-state anisotropy is about 0.15, and this categorizes the compound as Case I.

\section{\texorpdfstring{$\textbf{Ce$_3$Al$_{11}$}$ }{Ce3Al11}}\label{Ce3Al11_appdix}
Single-crystal magnetization data for orthorhombic Ce$_3$Al$_{11}$ is available from Ref.~\cite{s.garde2008Electrical}, and it shows an easy-axis ordering along the $b$ axis. Both the inverse magnetic susceptibility and field-dependent magnetization in Ref.~\cite{s.garde2008Electrical} display a crossing in $a$ and $c$ axes. However, this crossing was not observed in the field-dependent magnetization data reported in Ref.~\cite{boucherle1995Magnetic}. This compound exhibits an FM ordering at $T_{C}$ = 6.3\,K, and an AFM ordering at $T_N$ = 4.5\,K. The CEF analysis suggests a $\lambda_b=-9$\,mol/e.m.u., $\lambda_a=-5$\,mol/e.m.u., and $\lambda_c=-20$\,mol/e.m.u. The fact that all $\lambda_i$ values are negative is likely related to the antiferromagnetic transition at $T_N = 4.5$\,K.

%%%%%%%%%%
\section{\texorpdfstring{$\textbf{CeNiSb$_2$}$ }{CeNiSb2}}\label{CeNiSb2_appdix}

The CEF analysis on single-crystal CeNiSb$_2$ has been studied in Ref.~\cite{thamizhavel2003Anisotropic}, with CEF parameters of $B_2^0=7$\,K, $B_4^0=0.3$\,K, and $B_4^4=6$\,K. This leads to a ground state of $0.531\ket{\pm 5/2}-0.847\ket{\mp 3/2}$, favoring $ab$ plane. %However, we were not able to replicate these fitting results. 
Unfortunately, there is no $C_{4f}$ data at the higher temperature region to analyze the Schottky anomaly and thus cannot constrain the CEF parameters. Therefore, we use the molecular field contribution values extracted from Ref.~\cite{thamizhavel2003Anisotropic} in our analysis. This compound requires further study.

%%%%%%%%%%

%%%%%%%%%%%%%%%%%

\section{CePdSb}\label{CePdSb_appdix}

\begin{figure*}[!htb]
\center
\includegraphics[width=\linewidth]{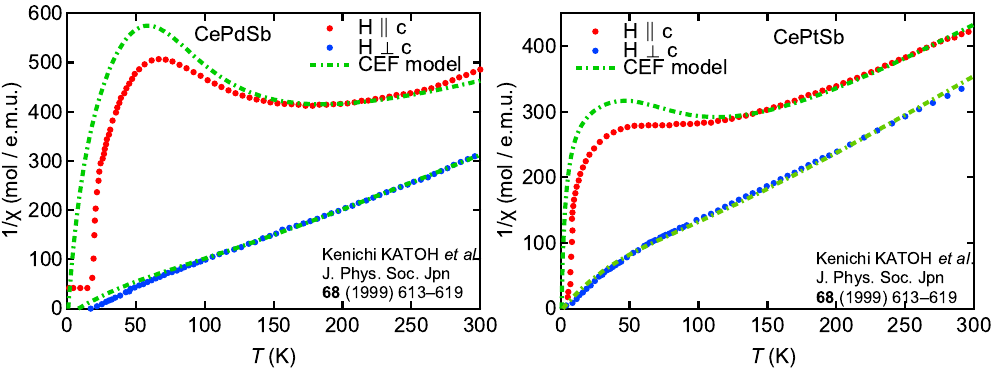}
\caption[]{Temperature-dependence of inverse magnetic susceptibility along the $c$ axis and $ab$ plane of trigonal compounds: (a) CePdSb[\ref{CePdSb_appdix}], and (b) CePtSb[\ref{CePtSb_appdix}]. The green dashed lines are the fitting based on the CEF model, and fitting parameters can be found in Table~\ref{CeFM_tri_table}.}
\label{CeFM_tri_figs}
\end{figure*}

\begin{table*}[]
\caption{Fitting parameters for the inverse magnetic susceptibility data on trigonal Ce-based compounds based on Eq.(A3) and Eq.(A4) in Ref.~\cite{jin2024Easyplane}.}
\begin{tabular}{l|l|l|l|p{0.10\linewidth}|p{0.10\linewidth}|p{0.10\linewidth}|p{0.10\linewidth}|l}
 & $B_{2}^{0}$ (K) & $B_{4}^{0}$ (K)   & $B_{4}^{3}$ (K)  & $\lambda_{\parallel}$ (mol/e.m.u) & $\lambda_{\perp}$ (mol/e.m.u) & $\chi_{0}^{\parallel} \times 10^{-5}$ (e.m.u./mol)          & $\chi_{0}^{\perp} \times 10^{-5}$ (e.m.u./mol)& Case\\ \hline

CePdSb~\cite{katoh1999Anisotropic,rainford1995Spin},[\ref{CePdSb_appdix}] & 15.77   &    -0.626   &  9.94  &  35.0 $\pm$ 22.8     &    14.4 $\pm$ 0.49    &    29.4 $\pm$ 10.3    &  4.82 $\pm$ 0.85 & (II)\\ 

CePtSb~\cite{katoh1999Anisotropic,rainford1994Structural},[\ref{CePtSb_appdix}] & 8.58   &    -0.545   &  14.8 &  45.6 $\pm$ 3.05     &    4.10 $\pm$ 0.74    &    1.62 $\pm$ 1.94    &  -2.02 $\pm$ 1.77 & (II)\\ 

\end{tabular}
\label{CeFM_tri_table}
\end{table*}

The single-crystal magnetization data for CePdSb were taken from Ref.~\cite{katoh1999Anisotropic}, as shown in Fig~\ref{CeFM_tri_figs}(a). A previous INS study reported the CEF parameters are $B_{2}^{0} = 15.77$\,K, $B_{4}^{0} = -0.626$\,K, and $B_{4}^{3} = 9.94$\,K~\cite{rainford1995Spin}. These parameters successfully reproduce the Schottky anomaly observed in heat capacity measurements~\cite{katoh2002Magnetic}. Using these CEF parameters, our CEF fitting yields $\lambda_{\parallel} = 35.0$\,mol/e.m.u. and $\lambda_{\perp} = 14.4$\,mol/e.m.u. The fitting parameters are listed in Table~\ref{CeFM_tri_table}. A previous study predicted that $\lambda_{\parallel} \approx$ 100\,mol/e.m.u. Our smaller value results from the inclusion of $\chi_{0}^{\parallel}$ to better fit the experimental data. This categorizes the compound as Case II.

\section{CePtSb}\label{CePtSb_appdix}
The single-crystal magnetization data for CePdSb were taken from Ref.~\cite{katoh1999Anisotropic}, as shown in Fig~\ref{CeFM_tri_figs}(b). A previous INS study reported the CEF parameters are $B_{2}^{0} = 8.58$\,K, $B_{4}^{0} = -0.545$\,K, and $B_{4}^{3} = 14.8$\,K~\cite{rainford1994Structural}. Using these CEF parameters, our CEF fitting yields $\lambda_{\parallel} = 45.6$\,mol/e.m.u. and $\lambda_{\perp} = 4.10$\,mol/e.m.u.  This categorizes the compound as Case II.

%%%%%%%%%%
\section{\texorpdfstring{$\textbf{CeRh$_6$Ge$_4$}$ }{CeRh6Ge4}}\label{CeRh6Ge4_appendix}

CeRh$_6$Ge$_4$ is an unusual FM KL that can be tuned to a quantum critical point by hydrostatic pressure~\cite{shen2020Strangemetal,kotegawa2019Indication}. The small magnetic entropy of $0.19R\ln2$ at $T_\textrm{C}$ and the $-\ln T$ dependence of the magnetic contribution to the electrical resistivity, suggests that the system is in close proximity to a quantum critical point~\cite{matsuoka2015Ferromagnetic}. The recent CEF analysis based on the inverse magnetic susceptibility has been studied in Ref.~\cite{shu2021Magnetic}. The CEF parameters were determined to be $B_2^0=1.25$\,meV (or $14.50$\,K), $B_4^0=0.0056$\,meV (or $0.065$\,K), together with $\lambda_{ab}=-52.8$\,mol/e.m.u. and $\lambda_{c}=-111$\,mol/e.m.u.~\cite{shu2021Magnetic}. The theoretical susceptibility expressions in the hexagonal point symmetry are:
\begin{equation} \label{eq:hex_chipara}
\chi_{B\parallel z}^{\textrm{total}} =\frac{N_{A}g_{J}^{2}\mu _\mathrm{{B}}^{2}\mu_{0}\beta}{Z}\Big[\frac{25}{4}+\frac{9}{4}(e^{-\beta \Delta_1})+\frac{1}{4}(e^{-\beta \Delta_2})\Big],
\end{equation}

\begin{equation}
\label{eq:hex_chiperp}
\begin{split}
\chi_{B\perp  z}^{\textrm{total}} =\frac{N_{A}g_{J}^{2}\mu _\mathrm{{B}}^{2}\mu_{0}}{k_{B}Z}
\Big[\frac{9e^{-\beta \Delta_2}}{4T} + 4(\frac{e^{-\beta \Delta_1}-e^{-\beta \Delta_2}}{\Delta_2-\Delta_1})\\+\frac{5}{2}(\frac{1-e^{-\beta \Delta_1}}{\Delta_1})\Big],
\end{split}
\end{equation}
where $\Delta_1 = -12B_{2}^{0}-240B_{4}^{0}$ and $\Delta_2 = -18B_{2}^{0}+60B_{4}^{0}$. The derivation can be found in Ref.~\cite{jin2022Suppression}. The expression can also be deduced by setting the $B_{4}^{4}$ and mixing angle $\alpha$ values in the tetragonal symmetry expressions [(Eq.~(\ref{eq:chipara_tetra}) and Eq.~(\ref{eq:chiperpen_tetra})] to zero.

\begin{figure}[]
\center
\includegraphics[width=\linewidth]{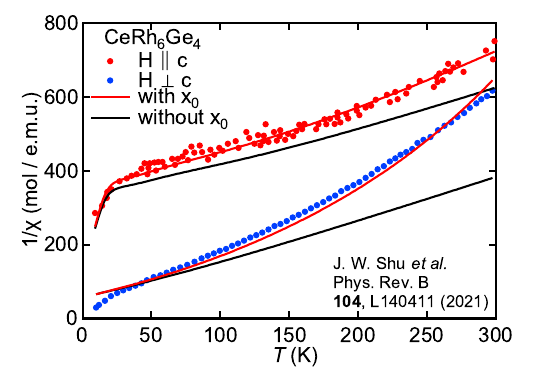}
\caption{Single-crystal magnetization data for CeRh$_6$Ge$_4$ were taken from Ref.~\cite{shu2021Magnetic}. A hexagonal CEF model including $\chi_0$ contributions was used to fit the published data, successfully replicating the previously reported CEF fitting results as shown in the red curves. For comparison, the same model without $\chi_0$ contributions was also applied as the black curves, highlighting the significant differences that arise when the $\chi_0$ value is large.}
\label{CeRh6Ge4_fig}
\end{figure}

We successfully replicated the published fitting results using the above expression with the reported CEF and molecular field parameters, as shown by the red curves in Fig.~\ref{CeRh6Ge4_fig}. Our results reveal $\chi_0^{\parallel}=-27.7\times 10^{-5}$ and $\chi_0^{\parallel}=-130.7\times 10^{-5}$ (e.m.u./mol).  %The origin of such a large residual susceptibility along the $ab$ plane remains unclear. The polycrystalline report of $\chi_0$ based on the CW fit has only about $-34.2\times 10^{-5}$ e.m.u./mol~\cite{matsuoka2015Ferromagnetic}.
This significant residual susceptibility strongly affects the accuracy of the CEF fitting. For comparison, the same CEF fitting without the $\chi_0$ contributions is shown as the black curve in Fig.~\ref{CeRh6Ge4_fig}. Notably, such a high $\chi_0^{\parallel}$ results in a difference of 265\,mol/e.m.u. at 300\,K. %Therefore, the published $\lambda$ values are not accurate. 
Additionally, there are no available $C_{4f}$ data at higher temperatures to analyze the Schottky anomaly and constrain the CEF parameters.

%%%%%%%%%%%%%
\section{\texorpdfstring{$\textbf{CeCuGa$_3$}$ }{CeCuGa3} }\label{CeCuGa3_appendix}

The single-crystal magnetization and heat capacity data for CeCuGa$_3$ were taken from Ref.~\cite{joshi2012Magnetic}. We obtain different CEF fit results, so  %using their published CEF parameters. The CEF parameters provided in Ref.~\cite{joshi2012Magnetic} yield CEF levels of $\Delta_1 = 5.25$\,K and $\Delta_2 = 228.7$\,K, rather than the reported $\Delta_1 = 50$\,K. Additionally, we could not reproduce the published Schottky anomaly using their reported CEF excitation levels. These splitting levels result in a sharp peak below the transition temperature and lower values in the higher temperature range (shown as the black dashed line in Fig.~\ref{CeCuGa3_fig}(a)), inconsistent with the experimental data. We redo the CEF analysis based on their experimental data. Unfortunately, we were unable to produce a conclusive CEF analysis result based on the existing data. In this section, 
we provide a comprehensive analysis to explain these discrepancies.

\begin{figure*}[]
\center
\includegraphics[width=\linewidth]{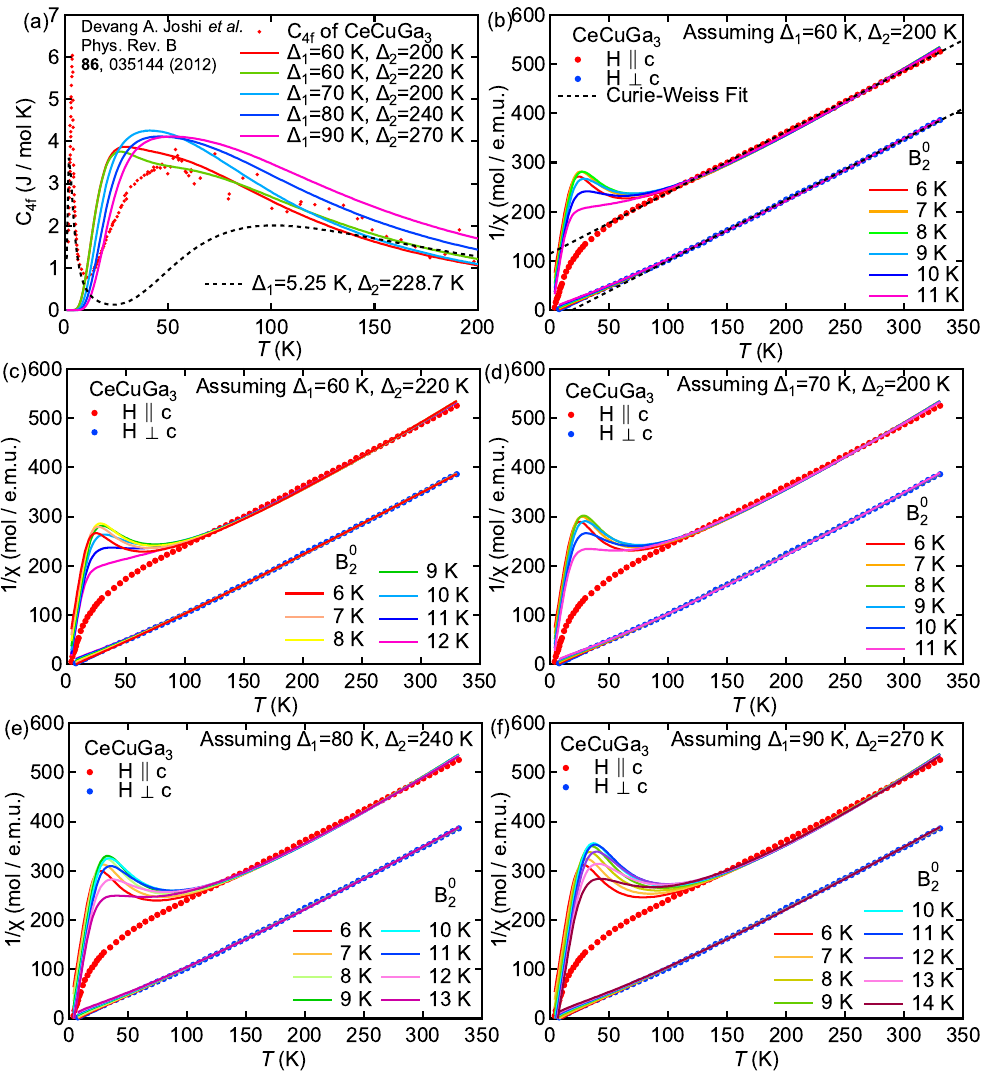}
\caption{(a) Heat capacity data for CeCuGa$_3$ were taken from Ref.~\cite{joshi2012Magnetic}. Five possible sets of theoretical Schottky anomalies that are close to experimental data are plotted. The black dashed line shows the correct Schottky anomaly if using the published CEF parameter in Ref.~\cite{joshi2012Magnetic}, which does not match well with the experimental data. The five possible sets of CEF splitting energies are (b)$\Delta_1=60$\,K, $\Delta_2=200$\,K, (c)$\Delta_1=60$\,K, $\Delta_2=220$\,K, (d)$\Delta_1=70$\,K, $\Delta_2=200$\,K, (e)$\Delta_1=80$\,K, $\Delta_2=240$\,K, and (f)$\Delta_1=90$\,K, $\Delta_2=270$\,K. The range $60-90$\,K in $\Delta_1$ was chosen so that the Schottky anomaly value will fit well near the lowest point in the experimental data around 11\,K. The corresponding CEF fittings using the incremental $B_2^0$ value (from 6\,K to the maximum $B_2^0$ value possible within the given constraints) in integer steps and respective CEF parameters are plotted in respective figures. The summary of all the fitting values is in Table~\ref{CeCuGa3_table}.}
\label{CeCuGa3_fig}
\end{figure*}

First, based on the field-dependent magnetization measurements, the saturated magnetization along the $ab$ plane is approximately 1.4\,$\mu_B$ at 7\,T~\cite{joshi2012Magnetic}. This value is significantly higher than 0.96$\mu_B$, indicating that the ground state must be $\Gamma_6 = \ket{\pm1/2}$ under the tetragonal symmetry. This behavior is similar to that observed in CeAuGa$_3$~\cite{lv2022Transport} and CeNi$_{0.74}$Ga$_{3.26}$~\cite{haddock2022Flux}. Thus, this will serve as the primary constraint for the CEF analysis. The heat capacity data for CeCuGa$_3$ is shown in Fig.~\ref{CeCuGa3_fig}(a). 

Unfortunately, no single set of CEF splitting energies fits the experimental data perfectly using the Eq.~\ref{HCSCH}. We identified five possible sets of CEF splitting energies that closely approximate the experimental data: (1)$\Delta_1=60$\,K, $\Delta_2=200$\,K, (2)$\Delta_1=60$\,K, $\Delta_2=220$\,K, (3)$\Delta_1=70$\,K, $\Delta_2=200$\,K, (4)$\Delta_1=80$\,K, $\Delta_2=240$\,K, and (5)$\Delta_1=90$\,K, $\Delta_2=270$\,K. The range of $\Delta_1=$60$-$90\,K was chosen so that the Schottky anomaly value will fit well near the lowest point in the experimental data around 11\,K. However, all of these sets yield Schottky anomaly values higher than the experimental data in the region between the transition temperature and 50\,K. New heat capacity measurements should be performed to confirm this discrepancy. %This discrepancy may arise from insufficient sample mass during the heat capacity measurements on CeCuGa$_3$ single crystals, which could lead to a low thermal coupling rate in the QD PPMS. To avoid this problem in our previous measurement in CeCuSi, we prepared larger masses of polycrystalline pellets by grinding selected single crystals from the same batch into powder and pressing them into pellets without further modification~\cite{jin2024Easyplane}.

The inverse magnetic susceptibilities of CeCuGa$_3$ along both axes are straight lines without a clear curvature signature at higher temperatures above 50\,K. These curvature signatures are often critical for accurately determining CEF parameters. In this case, we used the sets of CEF splitting energies mentioned earlier as constraints to fit the inverse susceptibility data. For each scenario, we assumed fixed CEF splitting energies and incremented the $B_2^0$ value in integer steps, starting from 6\,K and increasing to the maximum $B_2^0$ value possible within the given constraints.

With the $B_2^0$ values fixed, the corresponding $B_4^0$ and $B_4^4$ values can be calculated using Eq.~(\ref{tetra_delta_relation}). Using these values, the remaining fitting parameters were fitted to the experimental data for both orientations simultaneously above 100\,K, employing Eq.~(\ref{eq:inverchi}) with Eq.~(\ref{eq:chipara_tetra}) and Eq.~(\ref{eq:chiperpen_tetra}). All the fitting results are presented in Fig.~\ref{CeCuGa3_fig}(b)$-$(f), and the fitted CEF parameter values are summarized in Table~\ref{CeCuGa3_table}.

We note that all fittings in the inverse susceptibilities along the $ab$ plane (the easy-plane) are not distinguishable. The main difference between all these fittings is in the 20$-$80\,K region along the $c$ axis (the hard-axis).
In every scenario, higher $B_2^0$ values will result in larger $B_4^0$ and smaller $B_4^4$ values. This combination produces a reduced curvature behavior below 50\,K, bringing the fit closer to the experimental data. However, higher $B_2^0$ values also lead to much larger $\lambda_{\parallel}$ values, particularly when the constraint includes a higher $\Delta_2$ value. It remains unclear whether such high $\lambda$ values are physically meaningful. Additionally, as $B_2^0$ increases, $\lambda_{\parallel}$ transitions from a positive value to a small negative value, but the uncertainty in this region is of the same order or even larger than the magnitude of the negative value, making interpretation difficult.

The Curie-Weiss law fit, shown as black dashed lines in Fig.~\ref{CeCuGa3_fig}(b), yields effective moments and Curie-Weiss temperatures of 2.54\,$\mu_B$ and $-92.8$\,K along the $c$ axis, and are 2.54\,$\mu_B$ and $+18.7$\,K along the $ab$ plane, respectively. Using Eq.~(\ref{B20CW}), the expected $B_2^0$ value is approximately 11.6\,K, which places $\lambda_{\perp}$ in an ambiguous range, making it difficult to determine whether the interaction is FM-like or AFM-like.

As a result, we cannot provide a conclusive CEF analysis based on the current data. Additional measurements, such as soft x-ray absorption, inelastic neutron scattering, Raman spectroscopy, and additional heat capacity data, are necessary for a more definitive analysis.

\begin{table*}[]
\caption{Fitting parameters to the inverse magnetic susceptibility data of CeCuGa$_3$ based on Eq.~(\ref{eq:chipara_tetra}) and Eq.~(\ref{eq:chiperpen_tetra}). The first three CEF parameters are constrained by the sets of CEF splitting energies. More explanation of the fitting can be found in the main text. The corresponding fitting results are shown in Fig.~\ref{CeCuGa3_fig}.}

\begin{tabular}{|c|c|c|c|c|c|c|}
\bottomrule
 $B_{2}^{0}$ (K) & $B_{4}^{0}$ (K)   & $B_{4}^{4}$ (K)  & $\lambda_{\parallel}$ (mol/e.m.u.) & $\lambda_{\perp}$ (mol/e.m.u.) & $\chi_{0}^{\parallel} \times 10^{-5}$ (e.m.u./mol)          & $\chi_{0}^{\perp} \times 10^{-5}$ (e.m.u./mol) \\ \hline

\rowcolor{LightCyan}\multicolumn{7}{ |c| }{\textbf{(1) $\Delta_1=60$\,K, $\Delta_2=200$\,K, Fig.~\ref{CeCuGa3_fig} (b) }} \\ \hline
   6 & -0.322 & 2.607 & -18.11 $\pm$ 2.59 &  9.95 $\pm$ 2.14 & -18.86 $\pm$ 1.73 & -3.54 $\pm$ 2.63 \\
   7 & -0.256 & 2.574 &  -6.23 $\pm$ 2.31 &  7.39 $\pm$ 1.85 & -19.28 $\pm$ 1.46 & -4.37 $\pm$ 2.31 \\
   8 & -0.189 & 2.432 &   4.56 $\pm$ 2.06 &  5.08 $\pm$ 1.60 & -19.29 $\pm$ 1.24 & -5.28 $\pm$ 2.03 \\
   9 & -0.122 & 2.158 &  13.83 $\pm$ 1.84 &  3.01 $\pm$ 1.39 & -18.77 $\pm$ 1.06 & -6.26 $\pm$ 1.79 \\
  10 & -0.056 & 1.690 &  21.23 $\pm$ 1.61 &  1.15 $\pm$ 1.20 & -17.66 $\pm$ 0.90 & -7.30 $\pm$ 1.56 \\
  11 &  0.011 & 0.713 &  26.03 $\pm$ 1.39 & -0.51 $\pm$ 1.01 & -15.77 $\pm$ 0.76 & -8.40 $\pm$ 1.33 \\\hline

\rowcolor{LightCyan}\multicolumn{7}{ |c| }{\textbf{(2) $\Delta_1=60$\,K, $\Delta_2=220$\,K, Fig.~\ref{CeCuGa3_fig} (c) }} \\ \hline
  6 & -0.378 & 2.961 & $-13.35 \pm 2.49$ & $11.23 \pm 2.04$ & $-19.95 \pm 1.63$ & $-3.93 \pm 2.49$ \\
  7 & -0.311 & 2.976 &  $-0.68 \pm 2.21$ &  $8.60 \pm 1.75$ & $-20.55 \pm 1.36$ & $-4.75 \pm 2.18$ \\
  8 & -0.245 & 2.899 &  $11.27 \pm 1.98$ &  $6.23 \pm 1.52$ & $-20.84 \pm 1.15$ & $-5.66 \pm 1.92$ \\
  9 & -0.178 & 2.722 &  $21.81 \pm 1.79$ &  $4.09 \pm 1.34$ & $-20.65 \pm 1.00$ & $-6.63 \pm 1.71$ \\
 10 & -0.111 & 2.422 &  $30.48 \pm 1.62$ &  $2.16 \pm 1.18$ & $-19.86 \pm 0.87$ & $-7.66 \pm 1.53$ \\
 11 & -0.045 & 1.945 &  $36.96 \pm 1.42$ &  $0.43 \pm 1.02$ & $-18.44 \pm 0.74$ & $-8.74 \pm 1.33$ \\
 12 &  0.022 & 1.073 &  $40.52 \pm 1.17$ & $-1.11 \pm 0.83$ & $-16.20 \pm 0.60$ & $-9.86 \pm 1.09$ \\\hline

\rowcolor{LightCyan}\multicolumn{7}{ |c| }{\textbf{(3) $\Delta_1=70$\,K, $\Delta_2=200$\,K, Fig.~\ref{CeCuGa3_fig} (d) }} \\ \hline 
   6 & -0.350 & 2.412 & -16.41 $\pm$ 2.57 &  9.75 $\pm$ 2.11 & -19.51 $\pm$ 1.70 & -3.54 $\pm$ 2.60 \\
   7 & -0.283 & 2.404 &  -4.20 $\pm$ 2.30 &  7.16 $\pm$ 1.83 & -20.01 $\pm$ 1.44 & -4.36 $\pm$ 2.29 \\
   8 & -0.217 & 2.279 &   7.18 $\pm$ 2.06 &  4.82 $\pm$ 1.59 & -20.16 $\pm$ 1.22 & -5.27 $\pm$ 2.02 \\
   9 & -0.150 & 2.017 &  17.03 $\pm$ 1.85 &  2.71 $\pm$ 1.40 & -19.79 $\pm$ 1.05 & -6.25 $\pm$ 1.80 \\
  10 & -0.083 & 1.548 &  24.92 $\pm$ 1.65 &  0.82 $\pm$ 1.22 & -18.80 $\pm$ 0.91 & -7.28 $\pm$ 1.58 \\
  11 & -0.017 & 0.423 &  30.55 $\pm$ 1.43 & -0.88 $\pm$ 1.03 & -17.14 $\pm$ 0.76 & -8.37 $\pm$ 1.36 \\\hline

\rowcolor{LightCyan}\multicolumn{7}{ |c| }{\textbf{(4) $\Delta_1=80$\,K, $\Delta_2=240$\,K, Fig.~\ref{CeCuGa3_fig} (e) }} \\ \hline 
6 & -0.489 & 2.859 & -4.86 $\pm$ 2.38 & 12.17 $\pm$ 1.92 & -22.40 $\pm$ 1.49 &  -4.30 $\pm$ 2.33 \\
   7 & -0.422 & 2.964 &  9.41 $\pm$ 2.16 &  9.44 $\pm$ 1.68 & -23.31 $\pm$ 1.27 &  -5.13 $\pm$ 2.07 \\
   8 & -0.356 & 2.975 & 23.32 $\pm$ 2.05 &  6.97 $\pm$ 1.54 & -24.01 $\pm$ 1.13 &  -6.03 $\pm$ 1.93 \\
   9 & -0.289 & 2.893 & 36.18 $\pm$ 2.05 &  4.71 $\pm$ 1.49 & -24.31 $\pm$ 1.07 &  -7.00 $\pm$ 1.90 \\
  10 & -0.222 & 2.710 & 47.53 $\pm$ 2.12 &  2.66 $\pm$ 1.50 & -24.13 $\pm$ 1.05 &  -8.02 $\pm$ 1.93 \\
  11 & -0.156 & 2.403 & 56.96 $\pm$ 2.17 &  0.81 $\pm$ 1.49 & -23.40 $\pm$ 1.04 &  -9.08 $\pm$ 1.95 \\
  12 & -0.089 & 1.914 & 63.61 $\pm$ 2.10 & -0.87 $\pm$ 1.42 & -21.94 $\pm$ 0.98 & -10.19 $\pm$ 1.87 \\
  13 & -0.022 & 1.003 & 67.18 $\pm$ 1.88 & -2.37 $\pm$ 1.26 & -19.71 $\pm$ 0.86 & -11.33 $\pm$ 1.67 \\\hline

\rowcolor{LightCyan}\multicolumn{7}{ |c| }{\textbf{(5) $\Delta_1=90$\,K, $\Delta_2=270$\,K, Fig.~\ref{CeCuGa3_fig} (f) }} \\ \hline 
6 & -0.600 & 3.074 &  4.65 $\pm$ 2.28 & 14.13 $\pm$ 1.82 & -24.63 $\pm$ 1.37 &  -4.83 $\pm$ 2.18 \\
   7 & -0.533 & 3.252 & 20.38 $\pm$ 2.16 & 11.32 $\pm$ 1.65 & -25.77 $\pm$ 1.21 &  -5.67 $\pm$ 2.01 \\
   8 & -0.467 & 3.341 & 36.13 $\pm$ 2.24 &  8.76 $\pm$ 1.64 & -26.76 $\pm$ 1.17 &  -6.59 $\pm$ 2.04 \\
   9 & -0.400 & 3.347 & 51.21 $\pm$ 2.51 &  6.41 $\pm$ 1.78 & -27.45 $\pm$ 1.23 &  -7.56 $\pm$ 2.23 \\
  10 & -0.333 & 3.270 & 65.18 $\pm$ 2.89 &  4.26 $\pm$ 1.97 & -27.75 $\pm$ 1.34 &  -8.58 $\pm$ 2.52 \\
  11 & -0.267 & 3.106 & 77.62 $\pm$ 3.28 &  2.31 $\pm$ 2.17 & -27.59 $\pm$ 1.44 &  -9.64 $\pm$ 2.80 \\
  12 & -0.200 & 2.837 & 87.59 $\pm$ 3.54 &  0.52 $\pm$ 2.29 & -26.82 $\pm$ 1.50 & -10.74 $\pm$ 2.98 \\
  13 & -0.133 & 2.431 & 94.67 $\pm$ 3.62 & -1.10 $\pm$ 2.29 & -25.39 $\pm$ 1.49 & -11.86 $\pm$ 3.02 \\
  14 & -0.067 & 1.797 & 98.67 $\pm$ 3.47 & -2.57 $\pm$ 2.17 & -23.28 $\pm$ 1.41 & -13.02 $\pm$ 2.89 \\\hline
  
\rowcolor{LightCyan}\multicolumn{7}{ |c| }{\textbf{(6) $\Delta_1=90$\,K, $\Delta_2=230$\,K, not shown }} \\ \hline 
6 & -0.489 & 2.468 &  -5.11 $\pm$ 2.42 &  11.4 $\pm$ 1.95 & -22.69 $\pm$ 1.52 &  -4.16 $\pm$ 2.38 \\
7 & -0.422 & 2.589 &   9.17 $\pm$ 2.21 &  8.67 $\pm$ 1.72 &   -23.6 $\pm$ 1.3 &  -4.99 $\pm$ 2.13 \\
8 & -0.355 & 2.601 &  22.97 $\pm$ 2.12 &  6.19 $\pm$ 1.59 & -24.26 $\pm$ 1.17 &   -5.89 $\pm$ 2.0 \\
9 & -0.289 & 2.507 &  35.92 $\pm$ 2.14 &  3.94 $\pm$ 1.55 & -24.58 $\pm$ 1.12 &  -6.86 $\pm$ 1.98 \\
10 & -0.222 & 2.294 &  47.22 $\pm$ 2.21 &   1.9 $\pm$ 1.56 &  -24.38 $\pm$ 1.1 &  -7.88 $\pm$ 2.02 \\
11 & -0.156 & 1.922 &   56.6 $\pm$ 2.26 &  0.04 $\pm$ 1.55 & -23.62 $\pm$ 1.08 &  -8.95 $\pm$ 2.03 \\
12 & -0.089 & 1.257 &  63.17 $\pm$ 2.18 & -1.63 $\pm$ 1.47 & -22.14 $\pm$ 1.02 & -10.06 $\pm$ 1.95 \\

\bottomrule
\end{tabular}
\label{CeCuGa3_table}
\end{table*}

%%%%%%%%
\section{\texorpdfstring{$\textbf{CeNi$_{0.74}$Ga$_{3.26}$}$ }{CeNi0.74Ga3.26}}\label{CeNiGa3_appendix}

\begin{figure*}[]
\center
\includegraphics[width=\linewidth]{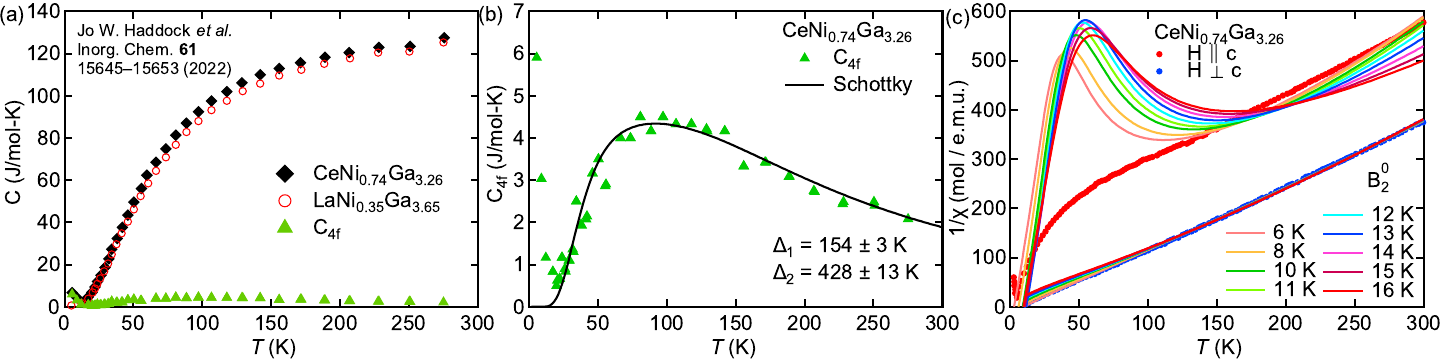}
\caption{(a) Heat capacity data for CeNi$_{0.74}$Ga$_{3.26}$ and LaNi$_{0.35}$Ga$_{3.65}$ were obtained from Ref.~\cite{haddock2022Flux}. The $4f$ contribution to the specific heat, $C_{4f}$, was calculated by subtracting the contributions of these two compositions. (b) The black line represents the Schottky anomaly fit to the $C_{4f}$, and yielding CEF splitting energies of $\Delta_1 = 154$\,K and $\Delta_2 = 428$\,K. (c) Using these CEF splitting energies, the inverse magnetic susceptibility data (from Ref.~\cite{haddock2022Flux}) in both orientations were fitted simultaneously above 100\,K. None of these parameters provides a satisfactory fit to the experimental data. The summary of all the fitting values is in Table~\ref{CeNiGa3_table}.}
\label{CeNiGa3_fig}
\end{figure*}

The single-crystal magnetization data of CeNi$_{0.74}$Ga$_{3.26}$ and heat capacity data of CeNi$_{0.74}$Ga$_{3.26}$ and LaNi$_{0.35}$Ga$_{3.65}$ were obtained from Ref.~\cite{haddock2022Flux}. First, we calculated the $4f$ contribution to the specific heat, $C_{4f}$, by subtracting the contributions of these two compositions, as shown in Fig.~\ref{CeNiGa3_fig}(a). A Schottky anomaly fit was performed and yielded CEF splitting energies of $\Delta_1 = 154$\,K and $\Delta_2 = 428$\,K. The saturated magnetization along the $ab$ plane is approximately 1.38\,$\mu_B$ at 7\,T~\cite{haddock2022Flux}. This value is significantly higher than 0.96$\mu_B$, indicating that the ground state must be the $\Gamma_6 = \ket{\pm1/2}$ state under the tetragonal symmetry. 

The Curie-Weiss temperatures for both axes are $\theta_{\parallel c}=-111$\,K and $\theta_{\perp c}=+16.3$\,K~\cite{haddock2022Flux}. Thus the expected $B_2^0$ value is approximately 13.6\,K. Therefore, a range of 6$-$16\,K for $B_2^0$ integer values were chosen, and $B_4^0$ and $B_4^4$ were then calculated based on the above two constraints. Using these CEF parameters, the remaining fitting parameters were fitted to the experimental data for both orientations simultaneously above 100\,K, employing Eq.~(\ref{eq:inverchi}) with Eqs.~(\ref{eq:chipara_tetra}) and ~(\ref{eq:chiperpen_tetra}). All the fitting results are presented in Fig.~\ref{CeNiGa3_fig}(c), and the fitted CEF parameter values are summarized in Table~\ref{CeNiGa3_table}. None of these parameters provides a satisfactory fit to the experimental data.

We suspect that the difficulties with the CEF fitting may arise from the subtraction of the heat capacity data for CeNi$_{0.74}$Ga$_{3.26}$ and LaNi$_{0.35}$Ga$_{3.65}$, which might not accurately isolate the $C_{4f}$ contribution from CeNi$_{0.74}$Ga$_{3.26}$. This issue could be due to the relatively large difference in Ni content between the two compounds. Further investigation is necessary to provide a more reliable CEF analysis.

%Notably, in CeNi$_{0.74}$Ga$_{3.26}$, there appears to be an AFM ordering at low temperatures near $T_\textrm{C}$ that is observed only along the $c$ axis (the CEF hard-axis) (see Fig.5(c) in Ref.\cite{haddock2022Flux}). The low-temperature heat capacity data reveal only one single transition. If confirmed, this compound could represent a rare example of an anisotropic RKKY system where FM ordering occurs along the $ab$ plane, and AFM ordering is restricted to the $c$ axis, both taking place at the same transition temperature.

\begin{table*}[]
\caption{Fitting parameters to the inverse magnetic susceptibility data of CeNi$_{0.74}$Ga$_{3.26}$ based on Eq.~(\ref{eq:chipara_tetra}) and Eq.~(\ref{eq:chiperpen_tetra}), using the constraint of CEF splitting energies $\Delta_1=154$\,K, $\Delta_2=428$\,K. These values were deduced from the Schottky anomaly fitting shown in Fig.~\ref{CeNiGa3_fig}(b). The corresponding CEF fitting results are shown in Fig.~\ref{CeNiGa3_fig}(c). However, none of these parameters provides a satisfactory fit to the experimental data.}

\begin{tabular}{|c|c|c|c|c|c|c|}
\bottomrule
 $B_{2}^{0}$ (K) & $B_{4}^{0}$ (K)   & $B_{4}^{4}$ (K)  & $\lambda_{\parallel}$ (mol/e.m.u.) & $\lambda_{\perp}$ (mol/e.m.u.) & $\chi_{0}^{\parallel} \times 10^{-5}$ (e.m.u./mol)          & $\chi_{0}^{\perp} \times 10^{-5}$ (e.m.u./mol) \\ \hline

   6 & -1.217 & 3.043 &   38.10 $\pm$ 4.72 &  15.18 $\pm$ 3.36 &  -56.57 $\pm$ 1.99 &  -19.67 $\pm$ 4.19 \\
   8 & -1.083 & 4.090 &   89.96 $\pm$ 7.36 &   9.45 $\pm$ 4.57 &  -57.77 $\pm$ 2.57 &  -22.58 $\pm$ 5.92 \\
  10 & -0.950 & 4.692 & 142.33 $\pm$ 11.79 &   4.45 $\pm$ 6.24 &  -57.33 $\pm$ 3.45 &  -25.57 $\pm$ 8.34 \\
  11 & -0.883 & 4.883 & 164.00 $\pm$ 14.89 &   2.19 $\pm$ 7.20 &  -55.53 $\pm$ 4.08 &  -27.08 $\pm$ 9.77 \\
  12 & -0.817 & 5.013 & 178.65 $\pm$ 18.69 &   0.08 $\pm$ 8.21 &  -51.94 $\pm$ 4.91 & -28.59 $\pm$ 11.31 \\
  13 & -0.750 & 5.086 & 182.76 $\pm$ 23.06 &  -1.91 $\pm$ 9.19 &  -46.03 $\pm$ 6.01 & -30.11 $\pm$ 12.81 \\
  14 & -0.683 & 5.105 & 174.94 $\pm$ 27.78 & -3.77 $\pm$ 10.04 &  -37.51 $\pm$ 7.47 & -31.62 $\pm$ 14.18 \\
  15 & -0.617 & 5.071 & 158.51 $\pm$ 32.49 & -5.50 $\pm$ 10.72 &  -27.05 $\pm$ 9.27 & -33.13 $\pm$ 15.32 \\
  16 & -0.550 & 4.982 & 141.24 $\pm$ 36.47 & -7.13 $\pm$ 11.18 & -16.78 $\pm$ 11.08 & -34.63 $\pm$ 16.14 \\

\bottomrule
\end{tabular}
\label{CeNiGa3_table}
\end{table*}
%%%%%%%%%%

%%%%%%%%%%%%%
\section{CeRuPO}\label{CeRuPO_appdix}

The single-crystal magnetization data for CeRuPO were taken from Ref.~\cite{krellner2008Single}. The previous Schottky anomaly analysis suggests that the first CEF gap is about 70\,K~\cite{krellner2007CeRuPO}, and the saturation moment suggests that the ground state is $\Gamma_6$~\cite{krellner2008Single}. With these restrictions, the proposed CEF parameters are listed in Table~\ref{CeFM_tetra_table}, and the fitting curves are shown in Fig.~\ref{CeFM_tetra_figs}(c). This categorizes the compound as Case III. Note that the $\lambda_{\parallel}$ is hard to pinpoint because the $c$ axis susceptibility magnitude is much smaller and thus more sensitive to changes in $\chi_0$ during the fitting. But the $\lambda_{\perp}$ shows a consistent negative value when applying different restrictions, supporting the argument that there is an AFM interaction along the CEF easy-plane that leads to the hard-axis ordering.

%%%%%%%%%%
\section{Other Ce compounds in the list}\label{otherCe_appdix}

$\textbf{CeVSb$_3$:}$ The magnetization data of orthorhombic CeVSb$_3$ shows that it has easy-axis ordering~\cite{sefat2008Magnetization}. At the ground state, the magnetic moments along the $b$ axis are slightly stronger than those along the $a$ axis. We did not perform CEF analysis due to a lack of inverse magnetic susceptibility data over the full temperature range.

$\textbf{Ce$_7$Rh$_3$:}$ Single-crystal magnetization data for hexagonal $\beta$-CeNiSb$_3$ is available from Ref.~\cite{tsutaoka2008Ferromagnetic}, and it shows an easy-axis ordering. We did not perform CEF analysis because this compound has three inequivalent Ce sites.

$\textbf{CePtAl:}$ Single-crystal magnetization data for orthorhombic CePtAl is available from Ref.~\cite{kitazawa1997Magnetic}, and it shows an easy-axis ordering along the $a$ axis. The inverse magnetic susceptibility data shows a crossing of $b$ and $c$ axes, but not in field-dependent magnetization up to 30\,T~\cite{kitazawa1997Magnetic}. We did not perform CEF analysis because of a lack of data from Schottky anomaly or INS studies to determine the CEF energy scheme.

$\textbf{CePtAl$_4$Si$_2$:}$ Single-crystal magnetization data for tetragonal CePtAl$_4$Si$_2$ is available from Ref.~\cite{ghimire2014Investigation}, and it shows an easy-plane ordering. We did not perform CEF analysis because of a lack of data from Schottky anomaly or INS studies to determine the CEF energy scheme.

$\textbf{Ce$_2$Pd$_2$In:}$ Single-crystal magnetization data for orthorhombic Ce$_2$Pd$_2$In is available from Ref.~\cite{klicpera2016Magnetic}, and it shows an easy-axis ordering along the $c$ axis. This compound shows an AFM ordering at $T_N$ = 4.5\,K, followed by an FM ordering at $T_c$ = 4.1\,K. We did not perform CEF analysis for this compound due to the lack of an analytical expression for the CEF Hamiltonian in orthorhombic symmetry.
%%%%%
%%%%%

%%%%%%%%%%
$\textbf{CeAuGa$_3$:}$ Single-crystal magnetization data for tetragonal CeAuGa$_3$ is available from Ref.~\cite{lv2022Transport}, but there is a lack of $C_{4f}$ data at higher temperatures needed to analyze the Schottky anomaly. Therefore, we did not perform CEF analysis on CeAuGa$_3$. The field-dependent magnetization along the $ab$ plane shows a saturation moment of about 1.3\,$\mu_\mathrm{B}$, which is well above 0.96\,$\mu_\mathrm{B}$. This suggests that the ground state is likely $\Gamma_6$.

$\textbf{CeGa$_2$:}$ Single-crystal magnetization data for hexagonal CeGa$_2$ is available from Ref.~\cite{takahashi1988Multiple}, and it shows an easy-plane ordering. This compound shows three AFM ordering at $T_{N1}$ = 9.9\,K, $T_{N1}$ = 10.3\,K, $T_{N1}$ = 11.3\,K, and an FM ordering at $T_c$ = 8.4\,K. We did not perform CEF analysis because of a lack of data from Schottky anomaly or INS studies to determine the CEF energy scheme.

$\textbf{CePtAl$_4$Si$_2$:}$ Single-crystal magnetization data for tetragonal CePtAl$_4$Si$_2$ is available from Ref.~\cite{ghimire2014Investigation}, and it shows an easy-plane ordering. We did not perform CEF analysis because of a lack of data from Schottky anomaly or INS studies to determine the CEF energy scheme.

$\textbf{CeCr$_2$Si$_2$C}$~\cite{wang2023Observation}, $\textbf{CeAlSi}$~\cite{yang2021Noncollinear}, and $\textbf{CeAgAl$_3$}$~\cite{muranaka2007Thermodynamic} are also FM KLs that exhibit easy-plane ordering. For these three compounds, we did not found inverse magnetic susceptibility data over the entire temperature range to perform the CEF analysis. However, the available magnetization data indicate that these compounds have easy-plane ordering without any magnetization curve crossing.

$\textbf{Ce$_5$Pb$_3$O:}$ Single-crystal magnetization data for tetragonal Ce$_5$Pb$_3$O is available from Ref.~\cite{macaluso2004Structure}, and it shows an easy-plane ordering.  We did not perform CEF analysis because this compound has two inequivalent Ce sites, and a lack of available inverse magnetization data over the entire temperature range.

$\textbf{CeRh$_3$B$_2$:}$ CeRh$_3$B$_2$ exhibits an exceptionally high Curie temperature of $T_C$ = 115\,K, with a small saturation moment of about 0.4\,$\mu_B/$Ce~\cite{dhar1981Strong}. Single-crystal magnetization data were reported in Ref.~\cite{galatanu2003Unusual}. Although the itinerant magnetism was proposed at the beginning owing to the high $T_C$ and small $\mu_s$, various experimental results indicated the localized nature of 4$f$ electrons, and the mechanism is still under debate~\cite{dhar1981Strong,kubota2013Weak,amorese2023Orbital}. The INS~\cite{givord2007Crystal}, magnetic anisotropy~\cite{givord2007Ferromagnetism}, and x-ray absorption experiment~\cite{amorese2023Orbital} are suggesting that $J=7/2$ multiplets must also be considered in the CEF analysis. Thus, our CEF analysis method is not applicable to this compound.

$\textbf{CeIr$_3$B$_2$:}$ Single-crystal magnetization data for CeIr$_3$B$_2$ were reported in Ref.~\cite{kubota2013Weak}. CeIr$_3$B$_2$ exhibits an unusually high Curie temperature $T_C$ = 41\,K, with a very small ordered moment of 0.04\,$\mu_B/$Ce. This compound also undergoes a structural phase transition from monoclinic at low temperature to hexagonal at 395\,K. There is no comprehensive CEF study on this compound, and it is possible that the CEF scheme of $J=7/2$ should be considered, as for CeRh$_3$B$_2$.

%%%%%%%%%%%%%%%%%%
%%%%%%%%%%%%%%%%
$\textbf{CeFeAs$_{0.7}$P$_{0.3}$O:}$ Single-crystal magnetization data for CeFeAs$_{0.7}$P$_{0.3}$O can be found in Ref.~\cite{jesche2012Ferromagnetism}. We did not perform CEF analysis because of a lack of available inverse magnetization data over the entire temperature range.  

$\textbf{$\beta$-CeNiSb$_3$:}$ Single-crystal magnetization data for orthorhombic $\beta$-CeNiSb$_3$ is available from Ref.~\cite{thomas2007Discovery}. We did not perform CEF analysis because this compound has two inequivalent Ce sites. At room temperature, the magnetic moment favors the $b$ axis, followed by the $c$ axis and $a$ axis. Similarly, the saturation moment at 7\,T is highest along the $b$ axis. However, at the ground state, the moment favors the $c$ axis, indicating that this is a hard-axis ordering case.

%%%%%%%%%%

%%%%%%%%%%%%%%%%%%%%%%%%%%%%%%%%%%%%%%%%%%%%%%%%%%%%%%%%%%%%%%%%%%%%%%%%%%%%%
%\bibliography{test,biblio}

\end{document}